\documentclass[aps,prep,epsfile,nofootinbib]{revtex4}

\usepackage{graphics}
\usepackage{slashed}
\usepackage{amssymb}
\usepackage{lscape}
\usepackage{amsmath}
\DeclareMathOperator{\arccot}{arcCot}
\usepackage{rotating}
\usepackage{hyperref}
\usepackage{mathrsfs}
\usepackage{graphicx}
\usepackage{xcolor}

\begin{document}

\title{Seminal Electromagnetic fields from preinflation}

\author{$^{1}$ Daniela Magos\footnote{ dmagoscortes@mdp.edu.ar} and $^{1,2}$ Mauricio Bellini
\footnote{mbellini@mdp.edu.ar}}
\address{
$^1$ Instituto de Investigaciones F\'{\i}sicas de Mar del Plata - IFIMAR, \\
Consejo Nacional de Investigaciones Cient\'ificas y T\'ecnicas
- CONICET, Mar del Plata, Argentina.\\
$^2$ Departamento de F\'{\i}sica, Facultad de Ciencias Exactas y
Naturales, Universidad Nacional de Mar del Plata, Funes 3350, C.P. 7600, Mar del Plata, Argentina.\\}
\begin{abstract}

We investigate the geometric dynamics of the primordial electric and magnetic fields during the early stages of the universe by extending a recently introduced quantum algebra \cite{BMAS}. We work on an extended model of gravity that considers the boundary terms from the Einstein-Hilbert action as geometric quantum fluctuations of the spacetime. We propose that the extended Riemann manifold is generated by a new connection $\hat{\delta\Gamma}^{\mu}_{\alpha\beta}$. This connection contains geometric information about the fluctuations of gravitational and electromagnetic fields in the vacuum, which could have been crucial during the primordial stages of the universe's evolution.
We revisit a preinflationary cosmological model \cite{mb} with a variable time scale and  negative spatial curvature, such that the universe begins with a null initial background energy density.
We observed the emergence of large scale magnetic fields starting from small values during the early phases of the universe's evolution. Subsequently, these fields decrease to reach present day values on the order of $\left<\hat{\delta B}\right> \simeq 10^{-12}\,{\rm G}$ on cosmological scales (between $10^{24}$ and $10^{26}$ meters). This significant deviation from inflationary models eliminates the need to impose excessively large initial values on these fields.
\end{abstract}
\maketitle

\section{Introduction and motivation}

One of the open questions in cosmology is to explain how the primordial electromagnetic fields and gravitational waves (GW) were generated and evolved in the primordial universe. The study of these fundamental phenomena is essential to gain insight into the physical processes that took place during the early expansion of the universe.
A particularly intriguing possibility is to provide a geometric explanation for the origin of electromagnetic field fluctuations. Their source is to be understood in the realm of quantum mechanics, and the expectation value of the flux associated with these fluctuations will locally modify the curvature of spacetime. Furthermore, it is possible to detect the primordial electromagnetic fields through the radiation of the cosmic microwave background, for instance, detecting traces of $B$-mode polarization in the electromagnetic waves would indirectly provide evidence for the existence of primordial GW. Concerning to the generation mechanisms of cosmological magnetic fields, it has been suggested that seed magnetic fields could originate from phase transitions occurring in the early universe \cite{6,7,thorne,rebhan,giova,rab,fammb}. However, it is natural to look for the possibility of generating large-scales magnetic fields during preinflation with strength according with observational data on cosmological scales: $\ll 10^{-9}$ G (Gauss). According to  \cite{mbs,prd0,prd}, the  origin of the magnetic fields could have been caused by certain primordial magnetic monopoles during preinflation. On intergalactic scales, there is evidence supporting the existence of a pre-galactic seed magnetic field on the order of $10^{-16}\,{\rm G}$. This is based on observations of high-energy TeV photons emitted by distant blazars \cite{26}. Studies conducted by \cite{27,66} suggest a present-day cosmological magnetic field strength of around $10^{-18} {\rm G}$.

However, the MAGIC collaboration \cite{MAGIC} has found that the intergalactic magnetic field from cosmological origin should be larger than $10^{-14}$ G. They studied the propagation of $\gamma$-rays in the intergalactic medium by analyzing the most recent 5-year observations from the MAGIC telescope, supplemented with data from the High Energy Stereoscopic System (H.E.S.S.), the Very Energetic Radiation Imaging Telescope Array System (VERITAS) and 12-years exposure of Fermi Large Area Telescope (LAT).

In \cite{Vela2023}, Da Vela {\it et al.} studied an extra galactic-$\gamma$ ray source to constrain the magnetic field, comparing their simulations with the results obtained by analyzing Fermi/LAT data. The Fermi/LAT and H.E.E.S.S collaborations \cite{ApJL950} update the lower limit of the intergalactic magnetic field to $>10^{-16}$ G. In a related study, \cite{Vovk2023} Vovk {\it et al.} derived a lower bound $10^{-19}$ G on the intergalactic magnetic field.

Inflation \cite{in0,in1,in11,in2,in3,in33,in333,in4,b1,bcms,npb} provides an optimal environment for the creation of large-scale primordial fields with significant coherence \cite{TW,Ra}. Preinflation \cite{Kur1,Kur2,Kur3,GR}, on the other hand describes an exponential \cite{MB}, (or super exponential \cite{dark23}) expansion with a null Hubble parameter initially, and later increases during the expansion. During preinflation the spatial curvature is negative and the universe is created without matter with initially null energy density. It is expected that during preinflation the large-scale coherence of primordial fields can be maintained and the rapid expansion during the era of super exponential expansion facilitates the generation of fields that are correlated on exceedingly large scales through the stretching of wave modes. Additionally, vacuum fluctuations of the electromagnetic field can be triggered while a mode is within the Hubble radius, and these can subsequently transform into classical fluctuations when modes exceed the Hubble radius \cite{mebe0}. Furthermore, it is expected that during preinflation, any existing densities of charged particles are greatly diluted by the expansion, resulting in the universe becoming a poor conductor on sub-Hubble scales. As a result, field generation from a zero field is not excluded by magnetic flux conservation. However, the Friedman-Robertson-Walker (FRW) universe is conformally flat, and the Maxwell theory exhibits conformal invariance, so magnetic fields generated in the preinflationary era would diminish significantly due to the accelerated expansion of the universe. One possibility to solve this problem relies on producing non-trivial magnetic fields in which conformal invariance is broken.

In this work, we shall introduce a connection given by,
\begin{equation}\label{conec}
\hat{\delta \Gamma}^{\mu}_{\alpha\beta}=b\,\hat{\Omega}^{\mu}\,g_{\alpha\beta},
\end{equation}
where $b$ is a parameter, $g_{\alpha\beta}$ is the metric tensor of the Riemannian manifold, and $\hat{\Omega}^{\mu}$ is a quantum field. The connection described by Eq. (\ref{conec}) encodes geometric information  concerning gravitational and electromagnetic fluctuations relative to the curved background represented by a Riemann manifold. It can be seen as a Weyl-type connection \cite{weyl}, incorporating both gravitational and electromagnetic aspects.

The manuscript is organized as follows: in Sect. II, we revisit an extend general relativity by using the connections proposed in the Eq. (\ref{conec}) to describe the boundary terms in the relativistic dynamics. Additionally, we derive the dynamics for the gravito-electromagnetic field and we obtain the dynamics for a particular gauge where both contributions, gravitational and electromagnetic are decoupled. In Sect. III we revisit the preinflationary model with variable time-scale and negative spatial curvature \cite{mb}, which we will use to study the evolution of electromagnetic fluctuations in the universe. In Sect. IV we develop the formalism for the quantum electromagnetic fluctuations in preinflation. We solve the dynamic equations, and in Sect. V we obtain the primordial electric and magnetic fields along with their corresponding expectation values during preinflation, using an extension of a new quantum algebra proposed in \cite{BMAS}. We also estimate the present-day values of co-moving magnetic fields on cosmological scales. Finally, in Sect. VI we present some final comments and conclusions.
\section{Extended General Relativity with boundary contributions}

Let us consider the Einstein-Hilbert action ${\cal I}$, which describes gravity and matter,
\begin{equation}\label{act}
{\cal I} =\int_V d^4x \,\sqrt{-g} \left[ \frac{R}{2\kappa} + {\cal L}_m\right],
\end{equation}
where $g$ is the determinant of the background metric tensor $g_{\alpha\beta}$, $R$ is the background scalar curvature, $\kappa = 8 \pi G/c^4$ is the Einstein gravitational constant ($G$ is the Newton constant of gravitation and $c$ is the light velocity in vacuum), and ${{\cal L}_m}$ is the Lagrangian density that describes the background physical dynamics. The variation of the Einstein-Hilbert action with boundary terms included is
\begin{equation}\label{delta0}
\delta {\cal I} = \int d^4 x \sqrt{-g} \left[ \delta g^{\alpha\beta}\left( G_{\alpha\beta} + \kappa T_{\alpha\beta}\right)
+ g^{\alpha\beta} \delta R_{\alpha\beta} \right]=0,
\end{equation}
where ${{T}}_{\alpha\beta}$ is the background stress tensor: ${{T}}_{\alpha\beta} =   2 \frac{\delta {{\cal L}_m}}{\delta g^{\alpha\beta}}  - g_{\alpha\beta} {{\cal L}_m}$. The Einstein tensor, denoted by $G_{\alpha\beta}=R_{\alpha\beta} - \frac{1}{2}\,R\,g_{\alpha\beta}$, is defined using the background (and symmetric) Ricci tensor $R_{\alpha\beta}$ which describes the geometry of the background using the Levi-Civita connections. The scalar curvature $R$ is given by $R= g^{\alpha\beta}\,R_{\alpha\beta}$.

To include the boundary terms in the global dynamics, we shall consider that the curvature variations $\delta R_{\alpha\beta}$ are originated by variations of the metric tensor \cite{RiBe,mobe}. In other words, we shall consider that $\delta g_{\alpha\beta}$ are the geometrical sources of $\delta R_{\alpha\beta}$,
\begin{equation}\label{f2}
\delta R_{\alpha\beta}= \lambda\left(x\right)\,\delta g_{\alpha\beta},
\end{equation}
where $\delta R_{\alpha\beta}$ and $\delta g_{\alpha\beta}$ will be considered on an extended manifold. Therefore the varied action, Eq. (\ref{delta0}), takes the form
\begin{equation}\label{del}
\delta {\cal I} = \int d^4 x \sqrt{-g} \left[ \delta g^{\alpha\beta} \left( G_{\alpha\beta} - \lambda\left(x\right) \,g_{\alpha\beta} + \kappa T_{\alpha\beta}\right)\right]=0.
\end{equation}
where we have used that  $\delta g^{\alpha\beta}\, g_{\alpha\beta} = - \delta g_{\alpha\beta}\, g^{\alpha\beta}$.
The Eq. (\ref{del}) implies that
\begin{equation}\label{tr}
G_{\alpha \beta} - \lambda\left(x\right) \,{g}_{\alpha \beta}=-\kappa\, T_{\alpha\beta}.
\end{equation}
To obtain the field equations, the boundary terms with $\lambda(x^\mu)$ in the Eq. (\ref{tr}) can be assimilated
to the redefined stress tensor: $\bar{T}_{\alpha\beta} = {T}_{\alpha\beta} - \frac{1}{\kappa} \lambda(x^\mu)\, g_{\alpha\beta}$, then we obtain that $\nabla_{\beta}\,G^{\alpha\beta}=\nabla_{\beta}\,\bar{T}^{\alpha\beta}=0$, and the background dynamics for the physical fields comes from the equations
\begin{equation}\label{v1}
\nabla_{\beta}\,T^{\alpha\beta} = \frac{1}{\kappa}\,g^{\alpha\beta}\,\frac{\partial\lambda(x^\mu)}{\hskip-.3cm\partial x^{\beta}}.
\end{equation}
Hence, the flux due to the boundary terms ($g^{\alpha\beta} \delta R_{\alpha\beta}$) in the minimized action, acts as the source for the physical fields.

\subsection{Quantum boundary terms}

In this work, we are interested in the flux originated from quantum gravitational and electromagnetic fields. This means that the varied connection, Eq. (\ref{conec}) will be quantum operators. In this case, the boundary terms in Eq. (\ref{delta0}), $ {g}^{\alpha \beta} \hat{\delta R}_{\alpha \beta}\equiv \hat{\delta\Theta}$, describe the flux of the 4-vector $\hat{\delta W}^{\alpha}=b^{-1}\left(\hat{\delta\Gamma}^{\epsilon}_{\beta\epsilon} {g}^{\beta\alpha}-\hat{\delta \Gamma}^{\alpha}_{\beta\gamma} {g}^{\beta\gamma}\right)$, through the 3D closed hypersurface $\partial M$; $\hat{\delta\Theta} = \nabla_\alpha\hat{\delta W}^\alpha$. Explicitly, the flux takes the form
\begin{equation}\label{FluxTh}
\hat{\delta\Theta} = \lambda(x^{\mu}) g^{\alpha\beta} \hat{\delta g}_{\alpha\beta} = -3\,\nabla_{\alpha}\,\hat{\Omega}^{\alpha},
\end{equation}
where $\hat{\Omega}^{\alpha}$ will be a 4-vector field that takes into account gravitational and electromagnetic contributions, that describes the geometric perturbations with respect to the background manifold. The differential operator $\nabla_{\alpha}$ denotes the covariant derivative on the Riemann manifold. To depict the variation of the Ricci tensor on the extended manifold, $\hat{\delta R}_{\alpha\beta}$, we shall use an extension of the Palatini expression \cite{pal}
\begin{equation}
\hat{\delta R}_{\alpha\beta} = b^{-1}\,\left[\left( \hat{\delta\Gamma}^{\mu}_{\alpha\mu} \right)_{\| \beta} - \left(\hat{\delta\Gamma}^{\mu}_{\alpha\beta}\right)_{\| \mu}\right].
\end{equation}
Here, $\left(...\right)_{\| \mu}$ denotes the covariant derivative of $\left(...\right)$ on the extended manifold. The covariant derivative of the background metric tensor on the extended manifold with self-interactions included \cite{mb}, is
\begin{equation}\label{cov}
\hat{g}_{\alpha\beta \|\mu} = \nabla_{\mu}g_{\alpha\beta} -\hat{\delta \Gamma}^{\nu }_{\alpha\mu}\,g_{\nu\beta} -\hat{\delta \Gamma}^{\nu }_{\beta\mu}\,g_{\alpha\nu} + 2\eta\,g_{\alpha\beta}\,\hat{\Omega}_{\mu},
\end{equation}
where $\nabla_{\mu}g_{\alpha\beta}=0$ is the covariant derivative of the metric tensor on the Riemann manifold and $\eta$ is a constant that considers the magnitude of self-interactions. In this framework, the variation of the metric tensor on the extended manifold, can be defined as
\begin{equation}\label{var}
\hat{\delta g}_{\alpha\beta} = \hat{g}_{\alpha\beta \|\mu}\,U^{\mu} = b\,\left( \hat{\Omega}^{\alpha}\, U^{\beta} + \hat{\Omega}^{\beta}\,U^{\alpha}\right) -
2\,\eta\,g^{\alpha\beta}\,\hat{\Omega}_{\mu}\, U^{\mu},
\end{equation}
where $U^{\mu}= \frac{d x^{\mu}}{\hskip -0.15cm d S}$ are the components of the relativistic observers that moves on the Riemann manifold.

By requiring the integrand in the Eq. (\ref{delta0}) to vanish, we obtain the following equation
\begin{equation}\label{gen}
\left[b\,\left(\hat{\Omega}^{\alpha} U^{\beta}+\hat{\Omega}^{\beta}\,U^{\alpha}\right) - 2\eta g^{\alpha\beta} \hat{\Omega}_{\mu}\,U^{\mu} \right]\,\left(G_{\alpha\beta}+\kappa\,T_{\alpha\beta}\right)= 3\left[\nabla_{\mu}\hat{\Omega}^{\mu} + \left(2b+\eta\right) \hat{\Omega}_{\mu}\,\hat{\Omega}^{\mu}\right].
\end{equation}

We are aimed to describe a linear equation for $\hat{\Omega}^{\mu}$, so that we use the gauge $\eta= - 2b$ into the Eq. (\ref{gen}), and using the Eq. (\ref{tr}), we obtain
\begin{equation}\label{uso}
\nabla_{\mu}\hat{\Omega}^{\mu} = 6\,b\,\lambda(x^{\alpha})\,\hat{\Omega}_{\mu} U^{\mu}.
\end{equation}
The way observers perceive the source varies based on their relative motion with respect to the source. This is reflected in the right side of the Eq. (\ref{uso}) by the components of the relativistic velocity, $U^{\mu}$. Additionally, we assume that the minimum spatial scale is the Planck length: $L_p=\left(\frac{\hbar G}{c^3}\right)^{1/2}$, which can be related directly with the parameter $b$, $L_p=\varepsilon \,b$, ($\varepsilon \ll 1$), such that, we can impose the following re-normalization
\begin{equation}\label{ufa}
\left<V\left|g^{\alpha\beta}\,\hat{\delta g}_{\alpha\beta}\right|V\right>=-\frac{1}{b},
\end{equation}
where $\left|V\right>$ is a quantum state on the Riemann manifold in the Heisenberg representation, meaning that the quantum operators evolve in spacetime while the quantum states are squeezed in time. Since our focus in this work is on cosmological applications, we will assume that the expectation value of a certain quantum operator $\hat{O}(x^{\mu})$ can be expressed as the space $3d$-volumetric expectation value $\left<V\right|\hat{O}(x^{\mu})\left|V\right>=O(t)$. Therefore the cosmological parameter will be given by
\begin{equation}\label{cp}
\lambda(t)=-b\,\left<V\left|\hat{\delta\Theta}\right|V\right> =3\,b\,\left<V\left|\nabla_{\mu}\hat{\Omega}^{\mu}\right|V\right> ,
\end{equation}
where $b=\frac{L_p}{\varepsilon}$  has units of length. When the metric tensor changes, it can cause fluctuations in the curvature of spacetime. Therefore, the curvature fluctuations can be expressed as the Eq. (\ref{f2}), where $\lambda(t)$ is proportional to the flux of $\hat{\Omega}^{\mu}$.

\subsection{Extended manifold for gravito-electromagnetic fields}

We shall propose the field $\hat{\Omega}^{\alpha}$ that incorporates both, gravitational and electromagnetic fluctuations in an unified description
\begin{equation}\label{omega}
\hat{\Omega}^{\alpha} \equiv \hat{\sigma}^{\alpha} + a\,\hat{\delta A}^{\alpha},
\end{equation}
where the trace of GW is
\begin{equation}\label{sig_psi}
    \hat\sigma \equiv g^{\alpha\beta} \hat{\delta\Psi}_{\alpha\beta}.
    \end{equation}
The gravitational perturbations are encoded in $\hat{\delta\Psi}_{\alpha\beta}$, and $a$ is a coupling constant with units $\left[a\right]={\rm \frac{s \,Coul}{kg\,m^3}}$. Furthermore $\hat{\sigma}_{\alpha}\equiv \hat{\sigma}_{,\alpha}$ denotes the partial derivative of $\hat{\sigma}$ with respect to the coordinate $x^{\alpha}$ and $\hat{\delta\hat{A}}^{\alpha}$ are the fluctuations of the electromagnetic $4$-potential.

Using Eqs. (\ref{omega}) and (\ref{sig_psi}) into the Eq. (\ref{gen}), we obtain the following equation of motion for the quantum fields
\begin{eqnarray}
&&\lambda(t)\,\left(2b-8\eta\right)\left(g^{\epsilon\theta}U^{\alpha}\nabla_{\alpha}\hat{\delta\Psi}_{\epsilon\theta}\,+a\,U_{\alpha}\,\hat{\delta A}^{\alpha}\,\right)- 3 (2b+\eta) g^{\epsilon\theta} g^{\alpha\beta} \nabla_{\mu} \hat{\delta\Psi}_{\epsilon\theta}\nabla^{\mu}\hat{\delta\Psi}_{\alpha\beta}\nonumber \\
&&-3\,g^{\epsilon\theta} \Box \hat{\delta \Psi}_{\epsilon\theta}- 3a\nabla_{\mu} \hat{\delta A}^{\mu}-6(2b+\eta)\,a\,g^{\epsilon\theta} \hat{\delta A}^{\mu}\,\nabla_{\mu}\hat{\delta\Psi}_{\epsilon\theta} \, - 3\left(2b+\eta\right)\,a^2 \hat{\delta A}^{\mu}\hat{\delta A}_{\mu}=0.
\label{gwef}
\end{eqnarray}
This is a non-linear differential equation for both, the GW's components $\hat{\delta\Psi}_{\epsilon\theta}$ and the fluctuations of the electromagnetic field $\hat{\delta A}^{\mu}$. Notice that both fields are coupled. Eq. (\ref{gwef}) is an important and exact result which is valid for arbitrary coupling constants: $a$, $b$ and $\eta$. Due to the difficulty of treating this equation in a general way, we shall
split Eq. (\ref{gwef}) in the following independent equations:
\begin{subequations}\label{18}
    \begin{equation}
        \Box \hat{\delta\Psi}_{\epsilon\theta}+ (2b+\eta) \nabla_{\mu} \hat{\delta\Psi}_{\epsilon\theta}\,\nabla^{\mu} \hat{\sigma} +2(2b+\eta)\,a\,\hat{\delta A}^{\mu}
\nabla_{\mu} \hat{\delta\Psi}_{\epsilon\theta} = \frac{2}{3}(b-4\eta) \,\lambda(t)\,U^{\mu}\,\nabla_{\mu} \hat{\delta\Psi}_{\epsilon\theta},
    \end{equation}
    \begin{equation}
        \nabla_{\mu} \hat{\delta A}^{\mu} + (2b+\eta) \,a\, \hat{\delta A}^{\mu} \hat{\delta A}_{\mu} = \frac{1}{3}(2b-8\eta) \,\lambda(t)\,U_{\mu}\hat{\delta A}^{\mu}.
    \end{equation}
\end{subequations}
Given that these equations are nonlinear and coupled, our next step will be to investigate a specific gauge that decouples the fields and linearizes the dynamical equations.

For $\eta=-2b$, the Eqs. (\ref{18}) can be written as two decoupled and linear differential equations
\begin{subequations}\label{ab}
    \begin{equation}\label{a}
        \Box \hat{\delta\Psi}_{\epsilon\theta} =6 \,b\,\lambda(t)\,U^{\mu}\,\nabla_{\mu} \hat{\delta\Psi}_{\epsilon\theta},
    \end{equation}
    \begin{equation} \label{b}
        \nabla_{\mu} \hat{\delta A}^{\mu}  = 6\,b\,\lambda(t)\,U_{\mu}\hat{\delta A}^{\mu}.
    \end{equation}
\end{subequations}
The Eq. (\ref{a}) describes the evolution of the GW's components $\hat{\delta\Psi}_{\epsilon\theta}$ with sources. The evolution for the trace of these components, $\hat{\sigma}=g^{\alpha\beta}\,\hat{\delta\Psi}_{\alpha\beta}$, can be written as
\begin{equation}
\Box \,\hat{\sigma} =6 b\,\lambda(t) \,U_{\mu} \,\hat{\sigma}^{\mu}.\label{c}
\end{equation}
The detailed study for the dynamics of the GW's in preinflation was presented in a previous work \cite{mb}.
On the other hand, the dynamics of the electromagnetic fluctuations $\hat{\delta A}^{\mu}$ during preinflation is described by Eq. (\ref{b}). In Sec. IV, we shall study analytical solutions of $\hat{\delta A}^{\mu}$, in a particular model of preinflation. Notice that when the system is isolated and the boundary terms in the Eq. (\ref{del}) are null, the standard Einstein equations with $\lambda(t)=0$ are recovered.

\section{Preinflation with negative spatial curvature}\label{preinf}

In this section, we will explore a model of the universe driven by a phantom scalar field, and electromagnetic fields minimally coupled to gravity. Assuming an electromagnetically neutral universe, the expectation value of these fields over the Riemannian background will be considered to be zero, only during the expansion, the fluctuations of the electromagnetic fields will be relevant. However, as we shall see in (\ref{cg}), the gravitational contributions of the geometric fluctuations will be applicable in the dynamics of the universe's expansion.

Let us study a Lagrangian density given by a scalar field $\hat{\varphi}$ and the electromagnetic fields encoded in the Faraday tensor $\hat F_{\mu\nu}$, to describe a preinflationary universe
\begin{equation}\label{lag}
{{\cal L}}_{m}= \left[\frac{1}{2} g^{\alpha\beta} \hat{\varphi}_{,\alpha}\hat{\varphi}_{,\beta} - V(\hat{\varphi})\right]-\frac{1}{4}\,\hat{F}^{\mu\nu}\,\hat{F}_{\mu\nu}.
\end{equation}
These fields are minimally coupled to gravity. The scalar field is responsible for the expansion of the universe and the electromagnetic tensor $\hat{F}_{\mu\nu}$ is defined by $\hat{F}_{\mu\nu}= \nabla_{\mu} \hat{A}_{\nu} - \nabla_{\nu} \hat{A}_{\mu}$,
where $\hat{A}^{\mu} \equiv \left( \hat\Phi(\vec{x},t)/c, \,\hat{A}^i(\vec{x},t)\right)$ is 4-electromagnetic potential such that $\hat{\Phi}(\vec{x},t)$ is the electric potential and $\hat{A}^i(\vec{x},t)$ is the vector potential. We assume an electromagnetically neutral universe on the background spacetime, such that
\begin{equation}\label{nul}
\left<V\right|\hat{A}^{\mu}(\vec{x},t)\left|V\right>=0.
\end{equation}
Therefore, the contributions of the electromagnetic potential to the dynamics will be given by the fluctuations $\hat{\delta A}^{\mu}(\vec{x},t)$.

To describe the background expanding universe with a variable time scale, we examine the line element where the spatial contributions are written in spherical coordinates
\begin{equation}\label{m1}
ds^2 = e^{-2\int \gamma(t)\,dt}\, c^2\,dt^2 - e^{2\int H(t)\,dt} \,\left[\frac{dr^2}{1-K\,r^2} + r^2\left(d\theta^2 + \sin^2(\theta) d\vartheta^2\right)\right].
\end{equation}
Here, $c$ is the light velocity in the vacuum, $H(t)=\dot a/a(t)$ is the Hubble parameter with a scale factor $a(t)$, $K=-(H_0/c)^2$ is the spatial curvature, with $H_0$ the Hubble parameter at certain time $t_0$,
and $\gamma(t)$ is a function that considers the variable timescale along the expansion of the universe \cite{rnbh}. Throughout the paper, we will be using
\begin{equation}\label{gamma}
\gamma\equiv-\frac{\dot{H}}{H},
\end{equation}
where the dot denotes derivative with respect to the time variable. The scalar field $\hat{\varphi}$, drives the expansion of the universe, and can be written as a semiclassical expansion \cite{bcms}
\begin{equation}
\hat{\varphi}\left(x^{\alpha}\right) = \left<V\right|\hat{\varphi}\left(x^{\alpha}\right)\left|V\right> + \hat{\delta\varphi}\left( x^{\alpha}\right).
\end{equation}
The expectation value of the scalar field $\hat{\varphi}\left(x^{\alpha}\right)$, is only a function of time, $\left<V\right|\hat{\varphi}\left(x^{\alpha}\right)\left|V\right>=\phi_c(t)$. The scalar field fluctuations are given by the differential equation
\begin{equation}
\ddot{\hat{\delta\varphi}} + \left(3H-\frac{\dot{H}}{H}\right)\,\dot{\hat{\delta\varphi}} - \left(\frac{H}{H_0}\right)^2\,e^{-2\int H(t)\,dt}\,\nabla^2 \hat{\delta\varphi} +  {\textcolor{orange}{\hat{\bar{\Upsilon}}}}''(\phi_c)\,\hat{\delta\varphi}=0,
\end{equation}
where
\begin{equation}\label{nabla}
\nabla^2\hat{\delta\varphi} \equiv \left(1-K\,r^2\right)\,\frac{\partial^2\hat{\delta\varphi}}{\partial r^2} + \frac{1}{r^2} \frac{\partial^2\hat{\delta\varphi}}{\partial \theta^2} + \frac{1}{r^2\,\sin^2(\theta)}
\frac{\partial^2\hat{\delta\varphi}}{\partial \vartheta^2} + \left( \frac{2}{r} - K r\right) \frac{\partial\hat{\delta\varphi}}{\partial r} + \frac{\cot{(\theta)}}{r^2}\,\frac{\partial \hat{\delta\varphi}}{\partial \theta}.
\end{equation}
The physical origin of the scalar field fluctuations $\hat{\delta\varphi}$ are responsible for the back-reaction effects that alter the background dynamics of the primordial universe \cite{dark231}, and must be consider to fully describe the dynamics of the preinflationary universe.

\subsection{Background dynamics and phantom field}

The dynamics of the expectation value for the scalar field, $\phi_c(t)$ is
\begin{equation}\label{pp}
\ddot{\phi}_c + \left[3\,H-\frac{\dot{H}}{H}\right] \dot\phi_c + \left.\frac{\delta \bar{\Upsilon}(\varphi)}{\delta\varphi}\right|_{\varphi\equiv\phi_c} =0,
\end{equation}
where the effective potential $\bar{\Upsilon}(\phi_c)$, with back-reaction contribution is
\begin{equation}\label{poo}
\bar{\Upsilon}(\phi_c)= \left[ V(\phi_c)\,\left(\frac{H}{H_0}\right)^2 + b\,\frac{\left<V\right|\hat{\delta\Theta}\left|V\right>}{\kappa}\right] = \left(\frac{H}{H_0}\right)^2\,\left[V(\phi_c)-\frac{\lambda(t)}{\kappa}\right].
\end{equation}
The extended Einstein equations with boundary terms included and $\gamma=-\dot{H}/H$, are given by
\begin{subequations}\label{ee}
    \begin{equation}\label{ee1}
        3 H(t)^2\left[ 1+ 3\,\alpha(t)\right]= \kappa \,\left[ \frac{\dot{\phi}^2_c}{2} +\bar{\Upsilon}(\phi_c)\right],
    \end{equation}
    \begin{equation}\label{ee2}
     -  H(t)^2\left[3+\alpha(t)\right] = \kappa\,\left[ \frac{\dot{\phi}^2_c}{2}- \bar{\Upsilon}(\phi_c)\right],
    \end{equation}
\end{subequations}
where the contribution of the spatial curvature is given by $\alpha(t)=(c^2\,K/H^2_0)\,e^{-2\int H(t)\,dt}$ \cite{mb}. Additionally, the total background pressure with boundary terms included takes the form $\bar{P}_T=\left(\frac{H}{H_0}\right)^2\,\left[ P+\frac{\lambda(t)}{\kappa}\right]$, and the total background energy density is
\begin{equation}\label{rho_b}
\bar{\rho}_T=\left(\frac{H}{H_0}\right)^2\,\left[\rho -\frac{\lambda(t)}{\kappa}\right].
\end{equation}
From the Eqs. (\ref{ee}), we obtain
\begin{eqnarray}
&& \dot{\phi_c}^2 = \frac{2\alpha(t)}{\kappa}\,H^2(t), \label{eee1}\\
&& \bar{\Upsilon}[\phi_c(t)] = \frac{[3+2\alpha(t)]}{\kappa} \,H^2(t). \label{eee2}
\end{eqnarray}
The equation of state is given by $\bar{\omega} = \frac{\bar{P}_T}{\bar{\rho}_T} = -1 + \frac{2\alpha}{3(1+\alpha)}$, which corresponds to
\begin{equation}
\phi_c(t) = \sqrt{\frac{2\alpha(t)}{\kappa}}.
\end{equation}
For $K<0$, we obtain that $\alpha(t) <0$ and $\phi_c(t)$ is an imaginary function. In this instance, the kinetic component of the energy density has a negative value, $\dot\phi_c^2/2<0$. This aligns with cosmological models featuring phantom fields that satisfy an equation of state: $\bar{\omega}<-1$. Similar to a previous study \cite{dark23}, we can adopt a Hubble parameter which becomes zero at $t=0$, and increases throughout the initial expansion. If we take $n\gg 1$, we can regard the Hubble parameter during preinflation as
\begin{equation}\label{hubb}
H(t)=H_0 \,\left(\frac{t}{t_0}\right)^{1/n},
\end{equation}
where we must remember that $H_0$ is the value of the Hubble parameter in some particular moment of the history of the universe: $t_0$. We shall choose $t_0$ at the moment when the Cosmic Microwave Background Radiation (CMBR) is emitted, $i.e$, at $t_0=3.8\times 10^5$ years. Furthermore, we will require that spatial curvature of the universe be negative $K=-(H_0/c)^2$, in order to the initial energy density value be null: $\left.\rho\right|_{t=0}=0$. This implies the hypothesis that initially, the universe is absent of matter in a background sense. Notice that the value for spacial curvature is the inverse of the squared Hubble distance at the moment when CMBR is emitted.

We are using $\gamma$ given by the Eq. (\ref{gamma}), such that, the metric of the Eq. (\ref{m1}) takes the form
\begin{equation}\label{m2}
ds^2 = \left(\frac{H(t)}{H_0}\right)^2\, c^2\,dt^2 - e^{2\int H(t)\,dt} \,\left[\frac{dr^2}{1+(H_0/c)^2\,r^2} + r^2\left(d\theta^2 + \sin^2(\theta)\, d\vartheta^2\right)\right],
\end{equation}
where $H(t)$ is the Hubble parameter of the universe given by the Eq. (\ref{hubb}). The choice of $g_{00}$ in the metric (\ref{m2}) is important because at $t=0$, the velocity $U_0$ will be null for a relativistic co-moving observer. In other words, the Eq. (\ref{m2}) describes an universe with variable time-scale, where the physical time $d\tau=\sqrt{g_{00}}\,dt$ starts running when the universe begins expanding.


\subsection{New algebra for the geometric fluctuations}\label{cg}

To compute the cosmological parameter $\lambda_0$, given by Eq. (\ref{cp})
it will be necessary to introduce a nonlinear quantum algebra associated with a differential operator $\hat{\frac{\partial}{\partial x^{\mu}}}$ \cite{BMAS}, providing a geometric invariant for the renormalization of the theory.
Using the definition of the field $\hat\Omega^\mu$ given in Eq. (\ref{omega}), we  obtain
\begin{equation}\label{expan}
\left<V\right|\hat{\Omega}^{\mu}\,U_{\mu}\left|V\right> = U_{\mu} \left<V\right|\hat{\sigma}^{\mu}\left|V\right> + a\, U_{\mu} \left<V\right|\hat{\delta A}^{\mu}\left|V\right>.
\end{equation}
Considering $\left<V\right|\hat{\delta A}^{\mu}\left|V\right>=0$, and using Eqs.  (\ref{uso}) and (\ref{cp}), we find that
\begin{equation}\label{AA}
\left<V\right|\hat{\Omega}^{\mu}\,U_{\mu}\left|V\right> = U_{\mu} \left<V\right|\hat{\sigma}^{\mu}\left|V\right>=\frac{1}{18\,b^2}.
\end{equation}
Moreover, we find that the cosmological parameter $\lambda_0$ will be purely of gravitational nature,
\begin{equation}\label{aa}
\lambda_0=-\,b\,\left<B\left|\hat{\delta\Theta}\right|B\right> =3\,b\,\left<B\left|\nabla_{\mu}{\hat{\sigma}}^{\mu}\right|B\right>,
\end{equation}
with $\hat{\delta\Theta}$ given by Eq. (\ref{FluxTh}).
Furthermore, from  the Eq. (\ref{b}), we obtain that
\begin{equation}
\left<V\right|\nabla_{\mu}\hat{\delta A}^{\mu}\left|V\right>=0.
\end{equation}
This result guarantees that the classical effects of the electromagnetic potential components on the background metric have a vanishing flux.

To calculate the expectation value of $\hat \sigma^{\mu}$, as given by Eq. (\ref{AA}), we must consider that the field $\hat{\sigma}$ is a quantum scalar field which can be written as a Fourier expansion in terms of their modes
\begin{equation}\label{sig1}
\hat{\sigma}\left(x^{\mu}\right) = \frac{1}{(2\pi)^{3/2}} \,\int d^3\,p\,\left[ \hat{C}_p\, {\sigma}(p,x^{\mu}) + \hat{C}^{\dagger}_p\,{\sigma}^*(p,x^{\mu})\right],
\end{equation}
where, $\hat{C}^{\dagger}_{p}$ and $\hat{C}_{p}$ are respectively the creation and annihilation operators. Furthermore, ${\sigma}(p,x^{\mu})$ corresponds to the modes of the field with momentum $p=\frac{k}{\hbar}$, where $k$ denotes the wavenumber.

To obtain the quantum representation of $\hat{\sigma}_{\alpha} \equiv \hat{\frac{\partial}{\partial x^{\alpha}}}\hat{\sigma}(x^{\mu})$, we use the nonlinear quantum algebra introduced in \cite{BMAS} to describe quantum differential operators:
\begin{equation}\label{dife}
\left[\hat{\frac{\partial}{\partial x'^{\alpha}}}, \hat{\sigma}^{\dagger}_p(x^{\mu})\right] = -\frac{i\,p'_{\alpha}}{\,\hbar } \frac{1}{ b^2}\,\left[\hat{\sigma}_{p'}(x'^{\mu}), \hat{\sigma}^{\dagger}_p(x^{\mu})\right],
\end{equation}
where $\hat{\sigma}^{\dagger}_p(x^{\mu})=\hat{C}^{\dagger}_p\,\sigma^*(p,x^{\mu})$ and $\hat{\sigma}_p(x^{\mu})=\hat{C}_p\,\sigma(p,x^{\mu})$. Furthermore, $p_{\alpha}$ are the $\alpha$th-components of the $4$-vector momentum ${\bf p}$. When the background spacetime is represented by orthogonal coordinates it is possible to develop the expansions
\begin{equation}
\sigma(p,x^{\mu})=e^{\frac{i}{\hbar} p_{\alpha}x^{\alpha}}\,\xi(p,\tau), \qquad \sigma^*(p,x^{\mu})=e^{-\frac{i}{\hbar} p_{\alpha}x^{\alpha}}\,\xi^*(p,\tau).
\end{equation}
Therefore, applying the algebra introduced in the Eq. (\ref{dife}) to the expansion (\ref{sig1}), we derive the expectation value of $\hat{\sigma}_{\alpha}$ on the background Riemann manifold
\begin{equation}\label{sin}
\left<V\right| \hat{\sigma}_{\alpha}\left|V\right> = \frac{-i}{(2\pi)^{3/2}\,\hbar^3 b^2} \int d^3p \left(\frac{p'_{\alpha}}{\hbar}\right)
\left<V \left| \left[\hat{C}_{p} ,\hat{C}^{\dagger}_{p'} \right]\right| V\right> \, \|\sigma(p,x^{\mu})\|^2,
\end{equation}
where
\begin{equation}\label{con}
\left<V\left| \left[\hat{C}_{p} ,\hat{C}^{\dagger}_{p'} \right]\right|V\right> = i \,\left(2\pi\right)^{3/2}\,\hbar^3 b^3\,\delta^{(3)} (\vec{p} - \vec{p}\,').
\end{equation}
Such that, the expectation value of $\hat{\sigma}_{\alpha}$, is
\begin{equation}\label{17}
\left<V\right| \hat{\sigma}_{\alpha}\left|V\right> = b\,\frac{p_{\alpha}}{\hbar}\, \left\| \sigma(p,x^{\mu}) \right\|^2,
\end{equation}
which allows obtaining the following invariant.
\begin{equation}\label{ufa}
\left<V\right| \hat{\sigma}_{\alpha}\left|V\right> \,U^{\alpha}=b\, \frac{p_{\alpha}\,U^{\alpha}}{\hbar}\, \left\| \sigma(p,x^{\mu}) \right\|^2=\frac{1}{18\,b^2},
\end{equation}
where we have used the Eq. (\ref{AA}). Thus, the introduction of the algebra in Eq. (\ref{dife}) allows for the possibility of a consistent theory of quantum gravity without divergences.

\subsection{Back-reaction effects}\label{bre}

In Sect. (\ref{cg}), we found that $\left< V \right|\nabla_{\mu}\hat{\delta A}^{\mu}\left|V \right>=0$, which implies that the cosmological parameter $\lambda_0$, as determined by Eq. (\ref{aa}), will be purely of gravitational nature. Will be pertinent to attribute the physical origin of $\lambda_0$ to $\hat{\delta\phi}=\hat\varphi -\left<\hat\varphi\right>$, because they describe the fluctuations related a massive quantum scalar field $\hat\varphi$
\begin{eqnarray}
\lambda_0 &=& \kappa \,\left<B\left|\frac{1}{2} g^{00}\,\left( \dot{\hat{\delta\phi}}\right)^2-
\frac{1}{2} g^{ij}\,\left(\nabla_{i}\hat{\delta\phi} \nabla_{j}\hat{\delta\phi}\right)+ \left.\sum_{n=1}^{\infty} \frac{1}{n!} \frac{\delta^{(n)} \textcolor{orange}{\hat{\bar{\Upsilon}}}(\hat{\varphi})}{\hat{\delta \varphi}^{(n)}}\right|_{\phi}\,\hat{\delta\phi}^{n}\right|B\right>  \nonumber \\
& = & \kappa\,\left<B\left|\hat{\rho}_T(\hat{\varphi})\right|B\right> - \kappa\,\bar{\rho}_T\left[\phi(t)\right] >0, \label{FFF}
\end{eqnarray}
such that $\left.\frac{\delta^{(n)} \textcolor{orange}{\hat{\bar{\Upsilon}}}(\hat{\varphi})}{\hat{\delta \varphi}^{(n)}}\right|_{\phi}$ is the $n$-th variation of the potential $V$ with respect to $\hat{\varphi}$, evaluated in the background field $\phi(t)$. From Eqs. (\ref{aa}) and (\ref{FFF}) we obtain that
the expectation value for the $\hat{\sigma}^{\alpha}$-flow will be related to the difference between the energy density contributions
\begin{equation}
\left<B\left|\nabla_{\mu}{\hat{\sigma}}^{\mu}\right|B\right> = \frac{\kappa}{3\,b}\,\Big(\left<B\left|\hat{\rho}_T(\hat{\varphi})\right|B\right> - \bar{\rho}_T\left[\phi(t)\right]\Big),
\end{equation}
which will be positive (negative) for $b>0$ ($b<0$). The equation (\ref{FFF}) means that the effects of boundary terms in the varied action (\ref{delta0}): $\left<B\left|{\hat{\delta\Theta}}\right|B\right>$ are sufficiently important to alter the evolution of the universe's background dynamics, because the scalar field fluctuations alter the global dynamics of the universe. This effect is represented by a cosmological parameter [see Eq. (\ref{FFF})], whose effects are proportional to $\left<B\left|\hat{\rho}_T(\hat{\varphi})\right|B\right> - \bar{\rho}_T\left[\phi(t)\right]$, where
the expectation value of the total energy density is given by
\begin{equation}
\left<B\left|\hat{\rho}_T(\hat{\varphi})\right|B\right> =  \left<B\left|\frac{1}{2} g^{00}\,\left( \dot{\hat{\varphi}}\right)^2-
\frac{1}{2} g^{ij}\,\left(\nabla_{i}\hat{\varphi} \nabla_{j}\hat{\varphi}\right)+\textcolor{orange}{\hat{\bar{\Upsilon}}}(\hat{\varphi})\right|B\right>,
\end{equation}
and $\bar{\rho}_T\left[\phi(t)\right]$ is the background energy density, given by Eq. (\ref{rho_b}).

\section{Primordial electromagnetic fluctuations}

To describe the electromagnetic fields we study a Lagrangian density ${\cal L}_{EM}$
\begin{equation}\label{lag}
{\cal L}_{EM} = -\frac{1}{4} \,\hat{F}^{\mu\nu}\,\hat{F}_{\mu\nu},
\end{equation}
The electromagnetic equations of motion, comes from the field equation $\nabla_{\mu} \hat{F}^{\mu\nu}=0$, that takes the form
\begin{equation}\label{wea}
\Box \hat{\delta A}^{\alpha} + g^{\alpha\nu}\,R_{\nu\mu}\,\hat{\delta A}^{\mu} =\nabla^{\alpha} \left(\nabla_{\mu} \hat{\delta A}^{\mu}\right) = 6\,\nabla^{\alpha} \left(\lambda(t)\,U_{0}\,\hat{\delta A}^{0}\right).
\end{equation}
In the last equality we are consider a co-moving frame with $U_0=\sqrt{g_{00}}$ and $U_i=0$.

The Eq. (\ref{wea}) describes the dynamics for the fluctuations of the electromagnetic field components, which fluctuate with respect to a null mean value, because we are assuming that the universe is electromagnetically neutral with respect to the background spacetime.

The four dynamical equations for $\hat{\delta A}^{\alpha}$, are explicitly
\begin{subequations}\label{eqmotion}
\begin{eqnarray}
\ddot{\hat \Phi}+\Big(3H-\gamma\Big)\dot{\hat\Phi}-\left(3H^2+3H \gamma +\dot\gamma\right)\hat\Phi - e^{-2\int(H+\gamma)dt}\bigg\{\nabla^2\hat{\Phi}\bigg. &&\nonumber\\
\bigg. -2H \left[\partial_1 \hat{\delta A}^1 + \partial_2\,\hat{\delta A}^2 + \partial_3 \hat{\delta A}^3 + \cot{(\theta)}\, \hat{\delta A}^2 + \left(\frac{2}{r}+ \frac{K\,r}{1-K\,r^2}\right)\hat{\delta A}^1\right] \bigg\} \,&=&0, \label{A0}\\
\ddot{\hat{\delta A}}^{1}+\Big(5H+\gamma\Big)\dot{\hat{\delta A}}^{1}+ 6H^2\,\hat{\delta A}^1-e^{-2\int(H+\gamma)dt}\left\{\nabla^2 \hat{\delta A}^1 + \frac{1}{1-Kr^2}\left(3K - \frac{1}{r^2}\right)\,\hat{\delta A}^1 \right.&&\nonumber\\ \left. +2 K \,r \,\partial_1 \hat{\delta A}^1
-2(1-K\,r^2)\left[h(t)\,\partial_1 \Phi-\frac{1}{r}\left(\partial_2 \hat{\delta A}^2+ \partial_3 \hat{\delta A}^3\right)\right]\right\}&=& 0,\label{A1}\\
\ddot{\hat{\delta A}}^{2}+\Big(5H+\gamma\Big)\dot{\hat{\delta A}}^{2}+ 6H^2\,\hat{\delta A}^2 - e^{-2\int(H+\gamma)dt}\left\{\nabla^2 \hat{\delta A}^2 -\left(4K - \frac{1}{r^2} + \frac{\cot^2{(\theta)}}{r^2}\right)\,\hat{\delta A}^2\right. &&\nonumber\\
\left.+ 2 \left(K \,r-\frac{1}{r} \right)\partial_1 \hat{\delta A}^2 -\frac{2}{r^2}\,\left[h(t)\,\partial_2 \Phi+\frac{1}{r}\partial_2 \hat{\delta A}^1- \cot{(\theta)}\partial_3 \hat{\delta A}^3\right]\right\}&=& 0,\label{A2}\\
\ddot{\hat{\delta A}}^{3}+\Big(5H+\gamma\Big)\dot{\hat{\delta A}}^{3}+ 6H^2\,\hat{\delta A}^3-e^{-2\int(H+\gamma)dt}\left\{\nabla^2 \hat{\delta A}^3 + 2\frac{\cot^2{(\theta)}}{r^2}\,\partial_2\hat{\delta A}^3 -4 K \,\hat{\delta A}^3\right.&&\nonumber\\
\left.- 2 \left(K \,r-\frac{1}{r} \right)\partial_1 \hat{\delta A}^3-\frac{2}{r^2 \sin^2{\theta}}\,\left[h(t)\,\partial_3 \Phi+\frac{1}{r}\partial_3 \hat{\delta A}^1+ \cot{(\theta)}\partial_3 \hat{\delta A}^2\right]\right\}&=& 0,  \label{A3}
\end{eqnarray}
\end{subequations}
where we have introduced the function
\begin{equation}\label{h}
h(t)\equiv H(t)-3\,b\,\lambda_0\,\left(\frac{H}{H_0}\right)^3,
\end{equation}
with $\lambda_0$ a constant given by the Eq. (\ref{aa}), and the Laplacian  $\nabla^2\hat{\delta A}^\alpha$, written in spherical coordinates, is
\begin{equation}\label{radial}
\nabla^2 \hat{\delta A}^\alpha \equiv \left(1+(H_0/c)^2\,r^2\right)\,\frac{\partial^2 \hat{\delta A}^\alpha}{\partial r^2}+ \frac{1}{r^2} \frac{\partial^2 \hat{\delta A}^\alpha}{\partial \theta^2} + \frac{1}{r^2\,\sin^2(\theta)}
\frac{\partial^2 \hat{\delta A}^\alpha}{\partial \vartheta^2} + \left( \frac{2}{r} + 3\,(H_0/c)^2 r\right) \frac{\partial \hat{\delta A}^\alpha}{\partial r} + \frac{\cot{(\theta)}}{r^2}\,\frac{\partial \hat{\delta A}^\alpha}{\partial \theta}.
\end{equation}
The spatial curvature $K=-(H_0/c)^2$ was chosen to achieve an initial null energy density. It is important to note that $H\gamma+\dot{H}=0$ due to Eq. (\ref{gamma}).

\subsection{Fluctuations of the Electric potential}

The dynamics for the fluctuations of the electric potential $\hat{\Phi}(x^{\alpha})$, are governed  by the Eq. (\ref{A0}), which, according to the conditions given in the Eq. (\ref{b}), takes the form of an homogeneous differential equation for $\hat{\Phi}$
\begin{equation}
\ddot{\hat \Phi}+\big(3H-\gamma+2h(t)\big)\dot{\hat\Phi}+\big(2H^2+2 h(t)\,(2H-3\gamma)+H\gamma-\dot\gamma\big)\hat\Phi - e^{-2\int(H+\gamma)dt}\nabla^2\hat{\Phi}=0,
\end{equation}
where $h(t)$ is given in the Eq. (\ref{h}).

Since we are interested in describing the electric potential during the preinflationary epoch, we can write it as a Fourier expansion in terms of their modes,
\begin{eqnarray}
\hat{\Phi}\left(x^{\alpha}\right)=\int
    \frac{d^3 k}{(2\pi)^{3/2}} \sum_{\iota=1}^{2}\,_{\iota} {\cal \varepsilon}^{0}(k)\,\,\sum_{l,m}\,\left[\varsigma^{(0)}(k,t) {}_\iota\hat\sigma_{klm}^{(0)}(x^j) +\left(\varsigma^{(0)}(k,t)\right)^* {}_\iota\hat\sigma_{klm}^{(0)\dagger}(x^j)\right],
\end{eqnarray}
where $-l\leq m \leq l$, ${}_\iota\hat\sigma_{klm}^{(0)}(x^j)={}_{\iota}\hat{A}_{klm}\, Y_{lm}(\theta,\vartheta)\, R^{(0)}_{l}\big(k\,(\vec{r}\cdot\hat{e}_r)\big)$ and ${}_\iota\hat\sigma_{klm}^{(0)\dagger}(x^j)= {}_{\iota}\hat A^{\dagger}_{klm}\, Y^*_{lm}(\theta,\vartheta)\, \left(R^{(0)}_{l}\big(k\,(\vec{r}\cdot\hat{e}_r)\big)\right)^*$. Here, $Y_{lm}(\theta,\vartheta)$ are the spherical harmonic and $R^{(0)}_{l}[k\,(\vec{r}\cdot\hat{e}_r)]$ are the solutions for the radial differential equation. The operators $_{\iota}\hat A^{\dagger}_{k}$ and $_{\iota}\hat A_{k}$ are the creation and destruction operators for a given $\iota$--polarization, that comply with the algebra
\begin{eqnarray}\label{m5}
&& \left[_{\iota}\hat A_{klm},\;_{\varsigma}\hat A_{k'l'm'}^{\dagger}\right]= \frac{(2l+1)}{4\pi} \frac{(l-m)!}{(l+m)!} \;\delta ({k}-{k'})\, \delta_{ll'}\,\delta_{mm'}\,\delta_{\iota\varsigma},\\
&& \left[_{\iota}\hat A_{klm},\;_{\iota}\hat A_{k'l'm'}\right]=
\left[ _{\iota}\hat A_{klm}^{\dagger},\;_{\iota}\hat A_{k'l'm'}^{\dagger}\right]=0,
\end{eqnarray}
for $_{\iota}\hat A_{klm}=\,_{\iota} \hat A_k\, A_{lm}$, such that $A_{lm}=\sqrt{\frac{(2l+1)}{4\pi} \frac{(l-m)!}{(l+m)!}}$. The indices $\iota$ denote the $2$-possible polarization modes and $\varsigma(k,t)$ are the time dependent modes for a given wave number $k$.


The canonical quantization requires that
\begin{equation}
\left[\hat{\delta A}_{0}(t,\vec{r}),\,\hat{\Pi}^{\mu}(t,\vec{r'})\right]= i \delta^{\mu}_0 \,\delta^{(3)}(\vec{r}-\vec{r'}),
\end{equation}
where $\hat{\Pi}^{\beta}=\frac{\delta {\cal L}_{EM}}{\hat{\delta A}_{\beta;\alpha}}\left.\right|_{\alpha=0}= g^{\mu\beta} \nabla_{\mu} \hat{\delta A}^{\alpha} - g^{\alpha\mu} \nabla_{\mu} \hat{\delta A}^{\beta}\left.\right|_{\alpha=0}$. To write the electric potential $\hat{\Phi}\left(x^{\alpha}\right)$ as real field, we can define the real operators $ _{\iota}\hat a_{klm}=\frac{1}{2} \left( _{\iota}\hat A_{klm}+ _{\iota}\hat A^{\dagger}_{klm}\right)$ and $\bar{R}^{(0)}_{l}[k\,(\vec{r}\cdot\hat{e}_r)]=\frac{1}{2}\left\{ {R}^{(0)}_{l}[k\,(\vec{r}\cdot\hat{e}_r)]+[{R}^{(0)}]^*_{l}[k\,(\vec{r}\cdot\hat{e}_r)]\right\}$
\begin{eqnarray}
\hat{\Phi}\left(x^{\alpha}\right)=\int
    \frac{d k}{2\pi^2} \sum_{\iota=1}^{2}\, {}_\iota{\cal P}^{0}({k},t)\,\,\sum_{l,m}\,k^2\,\left[ _{\iota}\hat a_{klm}\, \bar{Y}_{lm}(\theta,\vartheta)\, \bar{R}^{(0)}_{l}[k\,(\vec{r}\cdot\hat{e}_r)]
\right], \label{f}
\end{eqnarray}
where $\bar{Y}_{lm}(\theta,\vartheta)=\frac{1}{2}\left[{Y}_{lm}(\theta,\vartheta)+{Y}^*_{lm}(\theta,\vartheta)\right]$ are the real spherical harmonic. Furthermore, ${}_\iota {\cal P}^{0}({k},t)$ are the time-dependent polarization components of each $k$-mode, for the electric potential $ \hat{\Phi}\left(x^{\alpha}\right) \equiv c\;\hat{A}^0\left(x^{\alpha}\right)$.
\begin{eqnarray}\label{pol0}
_{1}{\cal P}^{0}= Re[\varsigma^{(0)}(k,t)], \quad _{2}{\cal P}^{0}= Im[\varsigma^{(0)}(k,t)].
\end{eqnarray}
The time dependent modes $\varsigma^{(0)}(k,t)$ obey the differential equation
\begin{equation}\label{varsigma}
    \ddot{\varsigma}^{(0)}+\big[3H-\gamma+2h(t)\big]\dot{\varsigma}^{(0)}+\Big[2H^2+2 h(t)\,\big(2H-3\gamma\big)+H\gamma-\dot\gamma +\kappa_0^2 \,e^{-2\int(H+\gamma)dt}\Big]\varsigma^{(0)} =0,
\end{equation}
where $\kappa_0$ is a constant to be determined.

\begin{figure}
    \centering
    \includegraphics[scale=0.55]{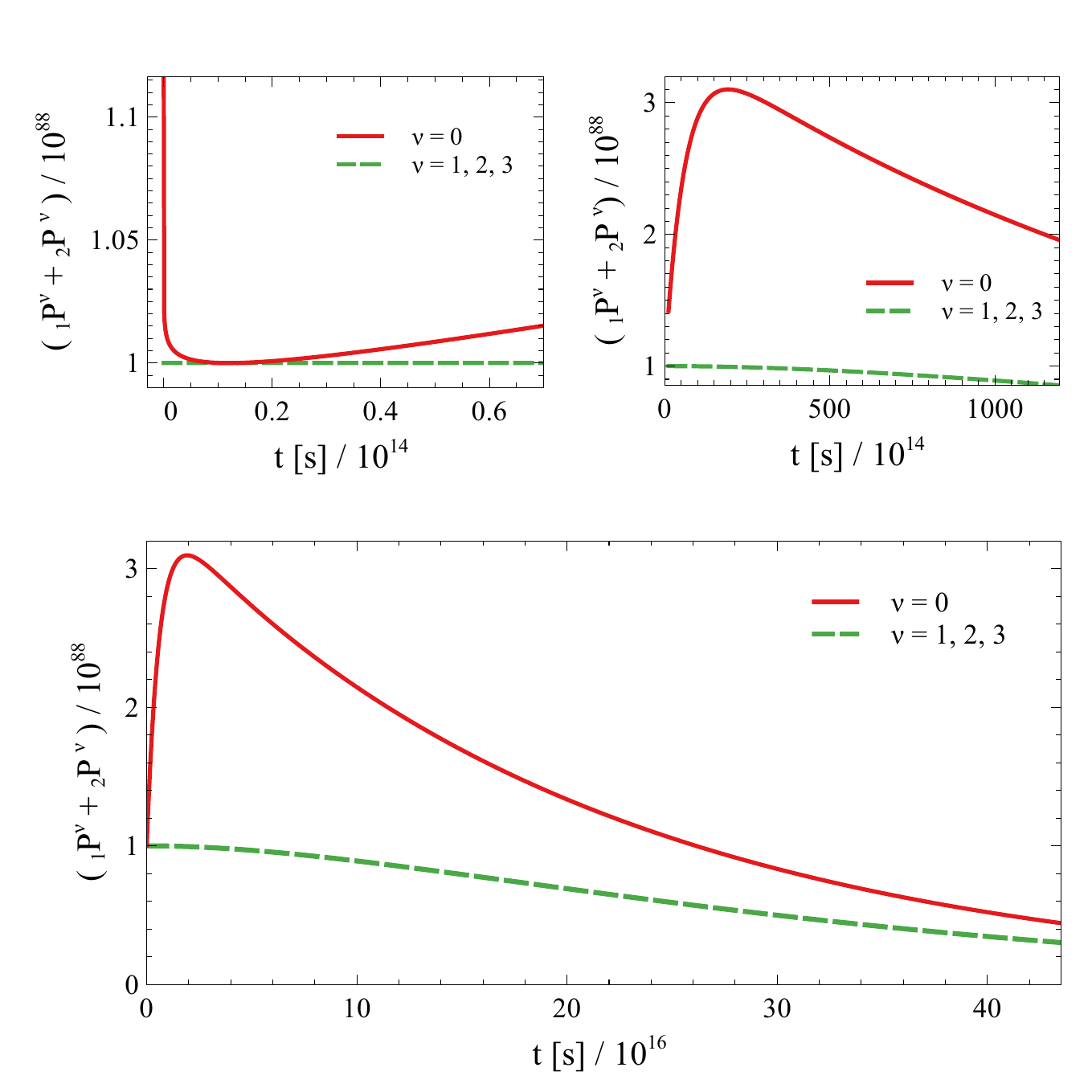}
    \caption{The polarization modes $\sum_{\iota} {}_\iota {\cal P}^{0}(k,t)$ of the electric potential $\hat{\Phi}$, (solid red line) and the polarization modes $\sum_{\iota} {}_\iota {\cal P}^{i}(k,t)$ of $\hat{\delta A}^{i}$ for $i=1,2,3$, (dashed green lines) are shown. All curves are depicted for $l=200$.}
    \label{fig:Polarization}
\end{figure}

In the Fig. \ref{fig:Polarization} we have plotted the sum of the polarization components for the electric potential $\hat{\Phi}$ (solid red line) with $l=200$. It is worth noting that in the early stages of the universe's evolution, the sum decreases from initially high values then begins to increase again until reaching its maximum shortly after the emission of the CMBR. Subsequently, it decreases monotonically until the present day.

The solution for the radial differential equation
\begin{eqnarray}
\big(1+(H_0/c)^2\,r^2\big) \frac{d^2}{d r^2}  \bar{R}^{(0)}_{l}(k\,r) + \left(\frac{2}{r} +3 (H_0/c)^2\,r\right) \frac{d }{ d r} \bar{R}^{(0)}_{l}(k\,r) +
\left(\kappa_0^2 - \frac{l(l+1)}{r^2} \right) \bar{R}^{(0)}_{l}(k\,r)=0, \label{l2}
\end{eqnarray}
is

\begin{eqnarray}
\bar{R}^{(0)}_l(k\,r)= \frac{1}{\sqrt{r}}&&\left\{{\rm C_1}\,{\it LP}\left[\sqrt{1-(c\,\kappa_0/H_0)^2}-1/2,\;\sqrt{l(l+1)+1/4}, \sqrt{1+(H_0/c)^2 r^2}\right] \right.\nonumber \\
 +&&  \;\; \left. {\rm C_2} \,{\it LQ}\left[\sqrt{1-(c\,\kappa_0/H_0)^2}-1/2,\;\sqrt{l(l+1)+1/4}, \sqrt{1+(H_0/c)^2 r^2}\right] \right\} , \label{solu}
\end{eqnarray}
where ${\rm C_1, C_2}$ are constants. Furthermore ${\it LP}\left[a_1,a_2,z\right]$ and ${\it LQ}\left[a_1,a_2,z\right]$ are the second kind associated Legendre functions, with parameters $a_1, a_2$ and variable $z$.

\subsection{Dynamics of $\hat{\delta A}^i$ components}

Similarly to the electric potential $\hat{\Phi}$, we propose a Fourier expansion for $\hat{\delta A}^i$, ($i=1,2,3$) as a complex field
\begin{eqnarray}\label{A:comp}
\hat{\delta A}^i\left(x^{\alpha}\right)=\int
    \frac{d^3 k}{(2\pi)^{3/2}} \sum_{\iota=1}^{2}\,_{\iota} {\cal \varepsilon}^{i}(k)\,\sum_{l,m}\,\left[\xi^{(i)}(k,t){}_\iota\hat{\sigma}_{klm}(x^j)+\,\left(\xi^{(i)} (k,t)\right)^*  {}_\iota\hat{\sigma}_{klm}^{(i)\dagger} (x^j)\right], \nonumber \\
\end{eqnarray}
where ${}_\iota\hat{\sigma}_{klm}^{(i)} (x^j)={}_{\iota}\hat B_{klm} \, Y_{lm}(\theta,\vartheta)\, R^{(i)}_{l}\big(k\,(\vec{r}\cdot\hat{e}_r)\big)$ and ${}_\iota\hat{\sigma}_{klm}^{(i)\dagger} (x^j) ={}_{\iota}B^{\dagger}_{klm}\, Y^*_{lm}(\theta,\vartheta)\, \left(R^{(i)}_{l} \big(k\,(\vec{r}\cdot\hat{e}_r)\big)\right)^*\,$, such that, the Eq.(\ref{A:comp}) can be written in terms of their modes as a real field,
\begin{eqnarray}
\hat{\delta A}^i\left(x^{\alpha}\right)=\int
    \frac{d k}{2\pi^2} \sum_{\iota=1}^{2}\,_{\iota} {\cal P}^{i}({k},t)\,\,\sum_{l,m}\,k^2\,\left( _{\iota}\hat b_{klm}^{(i)}\, \bar{Y}_{lm}(\theta,\vartheta)\, \bar{R}^{(i)}_{l}\left[k\,(\vec{r}\cdot\hat{e}_r)\right] \right), \label{ff}
\end{eqnarray}
where the functions $ \bar{Y}_{lm}(\theta,\vartheta)$ are the spherical harmonics and $ _{\iota}\hat b_{klm}^{(i)}=\frac{1}{2} \left( _{\iota}\hat B_{klm}^{}+ \,{}_\iota{\hat B_{klm}^{ \dagger}}\right)$, with $ _{\iota}\hat B_{klm}$ and $_{\iota}\hat B^{\dagger}_{klm}$ the destruction and creation operators, respectively, and they obey the algebra
\begin{eqnarray}\label{alg}
&& \left[_{\iota}\hat B_{klm},\;_{\varsigma}\hat B_{k'l'm'}^{\dagger}\right]=\frac{(2l+1)}{4\pi} \frac{(l-m)!}{(l+m)!} \; \delta ({k}-{k'})\, \delta_{ll'}\,\delta_{mm'}\,\delta_{\iota\varsigma},\\
&& \left[_{\iota}\hat B_{klm},\;_{\iota}\hat B_{k'l'm'}\right]=
\left[ _{\iota}\hat B_{klm}^{\dagger},\;_{\iota}\hat B_{k'l'm'}^{\dagger}\right]=0.
\end{eqnarray}
The polarization modes of $\hat{\delta A}^i$, are given by
\begin{eqnarray}\label{pol}
_{1}{\cal P}^{i}(k,t)= Re[\xi^{(i)}(\kappa_i,t)], \quad _{2}{\cal P}^{i}(k,t)=  Im[\xi^{(i)}(k,t)]
\end{eqnarray}
which define a plane which is orthogonal to the wave propagation.

\subsubsection{$r$-component $\hat{\delta A^1}$}

To solve the system of Eqs. (\ref{A1}-\ref{A3}), we can essay a particular solution of the Eq. (\ref{A1}) as two differential equations
\begin{subequations}
\begin{eqnarray}
 e^{2\int (H+\gamma)dt} \left[\ddot{\hat{\delta A}}^1 +\left(5H+\gamma\right) \dot{\hat{\delta A}}^1 + 6H^2\,\hat{\delta A}^1\right]
&=&\nabla^2 \hat{\delta A}^1 + \frac{2}{r}\partial_r \hat{\delta A}^1 +
\frac{3+3(H_0/c)^2\, r^2 +2 (H_0/c)^4\, r^4}{r^2\,\big(1+(H_0/c)^2\, r^2\big)} \,\hat{\delta A}^1,\label{aa1}  \\
-\cot{(\theta)}\,{\hat{\delta A}}^2 &=& \dot{\hat\Phi}+r\, h(t)\partial_r \hat\Phi+\left[2h(t)+H-\gamma\right]\hat\Phi
, \label{aa2}
\end{eqnarray}\end{subequations}
where $h(t)$ is given in the Eq. (\ref{h}) and we have used the condition provided by the Eq. (\ref{b}).

The time dependent modes $\xi^{(i)}(k,t)$ are solutions of the differential equation
\begin{equation}\label{time}
\ddot{\xi}^{(i)}+ \left[5H+\gamma\right]\,\dot{\xi}^{(i)} + \left[6H^2+{\kappa_i}^2\,e^{-2\int \left( H+\gamma \right) dt}
\right]\,{\xi^{(i)}}=0,
\end{equation}
for some constant $\kappa_i$ and $i=1$. The polarization components for fixed $k$ and a numeric solution for the Eq. (\ref{time}) is depicted in the Fig,  (\ref{fig:Polarization}) by the dashed green curve.
Furthermore, $\bar{R}^{(1)}_{l}\left[k\,(\vec{r}\cdot\hat{e}_r)\right]$ obeys the differential equation:

\begin{eqnarray}\label{R1}
\left(1+(H_0/c)^2\,r^2\right) \frac{d^2 }{d r^2}\bar{R}^{(1)}_{l}(k\,r)& +& \left(\frac{4}{r} +3 (H_0/c)^2\, r\right) \frac{d }{d r} \bar{R}^{(1)}_{l}(k\,r) \nonumber \\
&+&
\left[{\kappa_1}^2 +\frac{3+3(H_0/c)^2\, r^2 +2 (H_0/c)^4\, r^4}{r^2\,\big(1+(H_0/c)^2\, r^2\big)} - \frac{l(l+1)}{r^2} \right] \bar{R}^{(1)}_{l}(k\,r)=0. \label{rad1}
\end{eqnarray}
The general solution of the Eq. (\ref{R1}) can be written as follows
\begin{eqnarray}
\bar{R}^{(1)}_{l}(k\,r) = \frac{\sqrt{1+(H_0\,r/c)^2}}{r^{3/2}}\, &&\hspace{-.2cm}\left\{ {\rm C_1^{(1)}} {\it LP}\left[i\,\sqrt{1+(c\,\kappa_1/H_0)^2}-1/2,\;\sqrt{l(l+1)-3/4}, \sqrt{1+(H_0/c)^2 r^2}\right] \right.\nonumber \\
 +&&  \;\; \left. {\rm C_2^{(1)}} \,{\it LQ}\left[i\,\sqrt{1+(c\,\kappa_1/H_0)^2}-1/2,\;\sqrt{l(l+1)+3/4}, \sqrt{1+(H_0/c)^2 r^2}\right] \right\}
\end{eqnarray}
where ${\it LP}$  and $\it LQ$ are the associated Legendre functions, and $C_1^{(1)}$ and $C_2^{(1)}$ are constants.

On the other hand, from the Eq. (\ref{aa2}) we find that the angle $\theta$ must be constant, $\theta_*\equiv \arccot (\alpha_1)$ (for some $\alpha_1$), and in consequence the spherical harmonics only depends on $\vartheta$, and we can write
\begin{equation}\label{alpha2}
\big(\partial_\theta \bar{Y}_{lm}\big)\left.\right|_{\theta_*} = \alpha_2  \bar{Y}_{lm} (\theta_*, \vartheta).
\end{equation}
Besides, the following relations are satisfied
\begin{subequations}\begin{eqnarray}
    \alpha_1\,  _{\iota}\hat b_{klm}^{(2)} &=& - _{\iota}\hat a_{klm} \\
    _\iota {\cal P}^{2}(k,t)\bar{R}^{(2)}_{l}(k\,r) &=& _\iota {\cal \dot{P}}^{0}(k,t)\bar{R}^{(0)}_{l}(k\,r) + \big(2h(t)+H-\gamma\big) \, _\iota {\cal P}^{0}(k,t)\bar{R}^{(0)}_{l}(k\,r) + h(t) _\iota {\cal P}^{0} (k,t)r\frac{d\bar{R}^{(0)}_{l}}{\hspace{-.3cm}dr}.\label{P2R2}
\end{eqnarray}\end{subequations}

\subsubsection{$\theta$-component $\hat{\delta A^2}$}

To solve the Eq. (\ref{A2}), we proceed similarly to the previous case, and split it into
\begin{subequations}\begin{eqnarray}
   e^{2\int(H+\gamma)dt} \left[\ddot{\hat{\delta A}}^{2}+\Big(5H+\gamma\Big)\dot{\hat{\delta A}}^{2}+ 6H^2\,\hat{\delta A}^2\right] &=&\nabla^2 \hat{\delta A}^2 -\left(4K - \frac{1}{r^2} + \frac{\alpha_1^2}{r^2}\right)\,\hat{\delta A}^2+ 2 \left(K \,r-\frac{1}{r} \right)\partial_1 \hat{\delta A}^2 \label{A2a} \\
  \alpha_1\alpha_3 \hat{\delta A}^3&=& \alpha_2\left( h(t)\, \hat\Phi+\frac{1}{r}\hat{\delta A}^1\right),\label{A2b}
\end{eqnarray}\end{subequations}
from the Eq. (\ref{A2b}) we found that the angle $\vartheta$ is also a constant $\vartheta_* = \frac{1}{m} \arctan\left(\frac{\alpha_3}{m}\right)$. The equation for the time dependent modes $\xi^{(2)}(k,t)$ has the same form as in the Eq. (\ref{time}), with $i=2$. Such behaviour is showed in the Fig. \ref{fig:Polarization}.

The radial component $\bar{R}^{(2)}_{l}\left[k\,(\vec{r}\cdot\hat{e}_r)\right]$ satisfies the differential equation
\begin{eqnarray}
\left(1+\frac{H_0^2}{c^2}\,r^2\right) \frac{d^2 }{d r^2} \bar{R}^{(2)}_{l}(k\,r)+\frac{H_0^2}{c^2}\,r\frac{d}{ d r}\bar{R}^{(2)}_{l}(k\,r) +\frac{1}{r^2}
\left(1-4\frac{H_0^2}{c^2}r^2-\alpha_1^2-l(l+1)+{\kappa_2}^2 r^2 \right) \bar{R}^{(2)}_{l}(k\,r)=0. \label{rad2}
\end{eqnarray}
The general solution for the Eq. (\ref{rad2}), is
\begin{eqnarray}
\bar{R}^{(2)}_l(k\,r)= \sqrt{r}&&\left\{{\rm C_1^{(2)}}\,{\it LP}\left[i \sqrt{4+(c\,\kappa_2/H_0)^2}-1/2,\;\sqrt{l(l+1)-3/4+\alpha_1^2}, \sqrt{1+(H_0/c)^2 r^2}\right] \right.\nonumber \\
 +&&  \;\; \left. {\rm C_2^{(2)}} \,{\it LQ}\left[i\sqrt{4+(c\,\kappa_2/H_0)^2}-1/2,\;\sqrt{l(l+1)-3/4+\alpha_1^2}, \sqrt{1+(H_0/c)^2 r^2}\right] \right\}, \label{solR2}
\end{eqnarray}
where $C_1^{(2)}$ and $C_2^{(2)}$ are constants and $\it LP, \; LQ$ are the associated Legendre functions. Furthermore, regarding to the Eq.(\ref{A2b}), we can deduce the following relationships
\begin{subequations}
    \begin{eqnarray}
        \alpha_1\,\alpha_3 {}_{\iota}\hat b_{klm}^{(3)} &=& - _{\iota}\hat a_{klm}, \\
    _\iota {\cal P}^{3}(k,t)\bar{R}^{(3)}_{l}(k\,r) &=& h(t)\,  {}_\iota {\cal P}^{0}(k,t)\bar{R}^{(0)}_{l}(k\,r) + \alpha_2   {}_\iota {\cal P}^{1}(k,t)\frac{\bar{R}^{(1)}_{l}(k\,r)}{r},\\
    &=& \alpha_2  {}_\iota {\cal P}^{2}(k,t)\bar{R}^{(2)}_{l}(k\,r)+ \big(1-\alpha_2\big) h(t)\, {}_\iota {\cal P}^{0}(k,t)\bar{R}^{(0)}_{l}(k\,r). \label{A3toA0}
    \end{eqnarray}
\end{subequations}
Notice that from the Eqs. (\ref{P2R2}) and (\ref{A3toA0}) we can obtain $ _\iota {\cal P}^{3}(k,t)\bar{R}^{(3)}_{l}(k\,r)$, in terms of $\hat\Phi$.

\subsubsection{$\vartheta$-component $\hat{\delta A^3}$}

Following the same line of thought, we now proceed to break down the Eq. (\ref{A3}) into the following
\begin{subequations}\begin{eqnarray}
    e^{2\int(H+\gamma)dt}\left[\ddot{\hat{\delta A}}^{3}+\Big(5H+\gamma\Big)\dot{\hat{\delta A}}^{3}+ 6H^2\,\hat{\delta A}^3\right]&=&\nabla^2 \hat{\delta A}^3 + 2\frac{\alpha_1^2\, \alpha_2}{r^2}\,\hat{\delta A}^3 -4 K \,\hat{\delta A}^3
 - 2 \left(K \,r-\frac{1}{r} \right)\partial_1 \hat{\delta A}^3, \label{A3a}\\
- \frac{1}{r}\hat{\delta A}^1, &=& h(t)\,\hat\Phi+ \alpha_1\hat{\delta A}^2.\label{A3b}
\end{eqnarray}\end{subequations}
The time dependent modes $\xi^{(3)}(k,t)$ obeys the Eq. (\ref{time}), with $i=3$, and the corresponding polarization modes are depicted in Fig. \ref{fig:Polarization}.
Notice that the evolution of the polarization modes, ${}_\iota{\cal P}^{i}(k,t)$, for $\hat{\delta A}^{i}$, shown as dashed green lines in Fig. \ref{fig:Polarization}, are identical for $i=1,2,3$. However, they are smaller than the one for the electric potential, {\it i.e.,} $[{}_1{\cal P}^{i}(k,t) + {}_2{\cal P}^{i}(k,t)]\;<\;[{}_1{\cal P}^{0}(k,t) + {}_2{\cal P}^{0}(k,t)]$.

While the radial equation for $\bar{R}^{(3)}_{l}\left[k\,(\vec{r}\cdot\hat{e}_r)\right]$ satisfies the differential equation
\begin{eqnarray}
\left(1+\frac{H_0^2}{c^2}\,r^2\right) \frac{d^2 \bar{R}^{(3)}_{l}(k\,r)}{\hskip -1cm d r^2}+\left(\frac{4}{r} +5 \frac{H_0^2}{c^2}\,r\right) \frac{d \bar{R}^{(3)}_{l}(k\,r)}{\hskip -1.1cm d r}-\frac{1}{r^2}
\left[4\frac{H_0^2}{c^2}-2\alpha_1\alpha_2-l(l+1)-{\kappa_3}^2 r^2 \right] \bar{R}^{(3)}_{l}(k\,r)=0. \label{rad3}
\end{eqnarray}
The solution can be written in terms of associated Legendre functions $\it LP$ and $\it LQ$ as follows:
\begin{eqnarray}
\bar{R}^{(3)}_l(k\,r)= \frac{1}{r^{3/2}}&&\left\{{\rm C_1^{(3)}}\,{\it LP}\left[i (c\,\kappa_3/H_0)-1/2,\;\sqrt{l(l+1)+9/4-2 \alpha_1\alpha_2}, \sqrt{1+(H_0/c)^2 r^2}\right] \right.\nonumber \\
 +&&  \;\; \left. {\rm C_2^{(3)}} \,{\it LQ}\left[i(c\,\kappa_3/H_0)-1/2,\;\sqrt{l(l+1)+9/4-2 \alpha_1\alpha_2}, \sqrt{1+(H_0/c)^2 r^2}\right] \right\} , \label{solR2}
\end{eqnarray}
where $C_1^{(3)}$ and $ C_2^{(3)}$ are constants. Using the Eq. (\ref{A3b}) and the Fourier expansion proposed in the Eq. (\ref{ff}), we obtain that
\begin{subequations}
    \begin{eqnarray}
        {}_{\iota}\hat b_{klm}^{(1)} &=& _{\iota}\hat a_{klm}, \\
    _\iota {\cal P}^{1}(k,t)\bar{R}^{(1)}_{l}(k\,r) &=& r\, {}_\iota {\cal P}^{2}(k,t)\bar{R}^{(2)}_{l}(k\,r) -r\, h(t) {}_\iota {\cal P}^{0}(k,t)\bar{R}^{(0)}_{l}(k\,r).
    \end{eqnarray}
\end{subequations}

Summarizing, we have found that, for each pair of angles
\begin{equation}\label{fix_angles}
\big(\theta_*,\vartheta_*\big)= \left(\arccot{\left(\alpha_1\right)},\frac{1}{m}\arctan\left({\frac{\alpha_3}{m}}\right)\right),
\end{equation}
the spacial components of the electromagnetic 4-potential can be expressed in terms of the electric potential $\hat \Phi$, as follows
\begin{subequations}\label{vec_pot}
    \begin{eqnarray}
        \hat{\delta A}^1 &=& r \,\dot{\hat\Phi} + r^2\,h(t)\partial_r \hat\Phi + r\, \big[h(t) + H -\gamma\big] \hat \Phi,\\
        \hat{\delta A}^2 &=& -\frac{1}{\alpha_1} \bigg[\dot{\hat\Phi} + r\, h(t)\partial_r \hat \Phi + \big(2h(t)+H-\gamma\big)\hat\Phi \bigg], \\
        \hat{\delta A}^3 &=& -\frac{\alpha_2}{\alpha_1 \alpha_3} \bigg[\dot{\hat\Phi} + r\, h(t)\partial_r \hat \Phi + \big[h(t)+H-\gamma + h(t)/\alpha_2\big]\hat\Phi \bigg].
    \end{eqnarray}
\end{subequations}
Once obtained the $\hat{\delta A}^i$ fields, we can compute the primordial electric and magnetic fields in the early universe.
\section{Primordial electric and magnetic fields}

Let us consider the electromagnetic fields orthogonal to the relativistic velocity $U^{\mu}=\frac{dx^{\mu}}{dS}$, so that $U^{\mu} \,\hat{\delta E}_{\mu}=0$ and $U^{\mu}\,\hat{\delta B}_{\mu}=0$. Thus, we can express the electric and magnetic fields in a following compact form, \cite{review}
\begin{eqnarray}
\hat{\delta E}^{\mu} &=& \hat{F}^{\mu\nu}\,U_{\nu}, \\
\hat{\delta B}^{\mu} &=& \frac{1}{2} \,\epsilon^{\mu\nu\rho\lambda}\, U_{\nu}\, \hat{F}_{\rho\lambda}.
\end{eqnarray}
Here, $\epsilon^{\mu\nu\rho\lambda}$ is the totally anti symmetric Levy-Civita symbol, such that $\epsilon^{0123}=1,\; (-1)$ for any even (odd) permutation of (0,1,2,3) and 0 if any index is repeated. Furthermore, choosing a co-moving gauge, {\it i.e.}, $U^{0} = \sqrt{g^{00}}$ and $U^i=0$, the nonzero components of the electric and magnetic fields consist only of the following spatial components
\begin{subequations}
    \begin{eqnarray}
\hat{\delta E}^{i} &=& \hat{F}^{i0}\,U_{0}=  U_0\left(g^{ij} \hat\nabla_j \hat{\Phi} - g^{0j} \hat\nabla_j \hat{\delta A}^i\right),\label{dEi}\\
\hat{\delta B}^{i} &=& \frac{1}{2} \,\epsilon^{i0jk}\, U_{0}\, \hat{F}_{jk}= -\frac{1}{2} \, \epsilon^{0ijk}U_0\left(\hat\nabla_j (\hat{\delta A}_k) - \hat\nabla_k (\hat{\delta A}_j)\right), \label{dBi}
\end{eqnarray}
\end{subequations}

for $i=1,2,3$. With the metric given by Eq. (\ref{m2}), the components of the electric field, Eq. (\ref{dEi}), take the form
\begin{subequations}\label{electric}
    \begin{eqnarray}
    \hat{\delta E}^{1} &=& - (1+(H_0/c)^2\,r^2)\,\left(\frac{H(t)}{H_0}\right)\,e^{-\int 2H dt} \,\hat\partial_1\left(\hat{\Phi}/c\right) - \,\left(\frac{H(t)}{H_0}\right)^3\,\dot{\hat{\delta A}}^1 ,\\
     \hat{\delta E}^{2} &=& -\frac{1}{r^2}\,\left(\frac{H(t)}{H_0}\right)\,e^{-\int 2H dt} \,\hat\partial_2 \left(\hat{\Phi}/c\right) - \,\left(\frac{H(t)}{H_0}\right)^3\,\dot{\hat{\delta A}}^2,\\
     \hat{\delta E}^{3} &=& -\frac{1}{r^2 {\sin^2\theta}}\,\left(\frac{H(t)}{H_0}\right)\,e^{-\int 2H dt} \,\hat\partial_3\left(\hat{\Phi}/c\right) - \,\left(\frac{H(t)}{H_0}\right)^3\,\dot{\hat{\delta A}}^3 .\\
    \end{eqnarray}
\end{subequations}
On the other hand, the spatial magnetic field components, Eq. (\ref{dBi}), are
\begin{subequations}\label{mag}
    \begin{eqnarray}
        \hat{\delta B}^{1} &=& r^2\,\left(\frac{H(t)}{H_0}\right)\, e^{\int 2H dt}\left[ \tilde{\alpha_1}\, \hat\partial_2 \,\hat{\delta A}^3 - \hat\partial_3 \,\hat{\delta A}^2 +  2\alpha_1 \tilde{\alpha_1} \,\hat{\delta A}^3 \right],\\
        \hat{\delta B}^{2} &=& \,\left(\frac{H(t)}{H_0}\right)\,e^{\int 2H dt}\left[r^2 \,\tilde{\alpha_1} \,\hat\partial_1 \,\hat{\delta A}^3 - \frac{1}{1+(H_0/c)^2 \,r^2}\,\hat\partial_3 \,\hat{\delta A}^1 +  \frac{2}{r} \,\hat{\delta A}^3 \right],\\
        \hat{\delta B}^{3} &=& \,\left(\frac{H(t)}{H_0}\right)\,e^{\int 2H dt}\,\left[r^2 \,\hat\partial_1 \,\hat{\delta A}^2 - \frac{1}{1+(H_0/c)^2 \,r^2}\,\hat\partial_2 \,\hat{\delta A}^1 +  \frac{2}{r} \,\hat{\delta A}^2 \right],
    \end{eqnarray}
\end{subequations}
where $\tilde\alpha_1 \equiv \sin^2{\left[\arccot\left(\alpha_1\right)\right]}$, and the vector potential $\hat{\delta A}^i$ can be written as in the Eq. (\ref{A:comp}).
\begin{figure}
    \centering
    \includegraphics[scale=0.4]{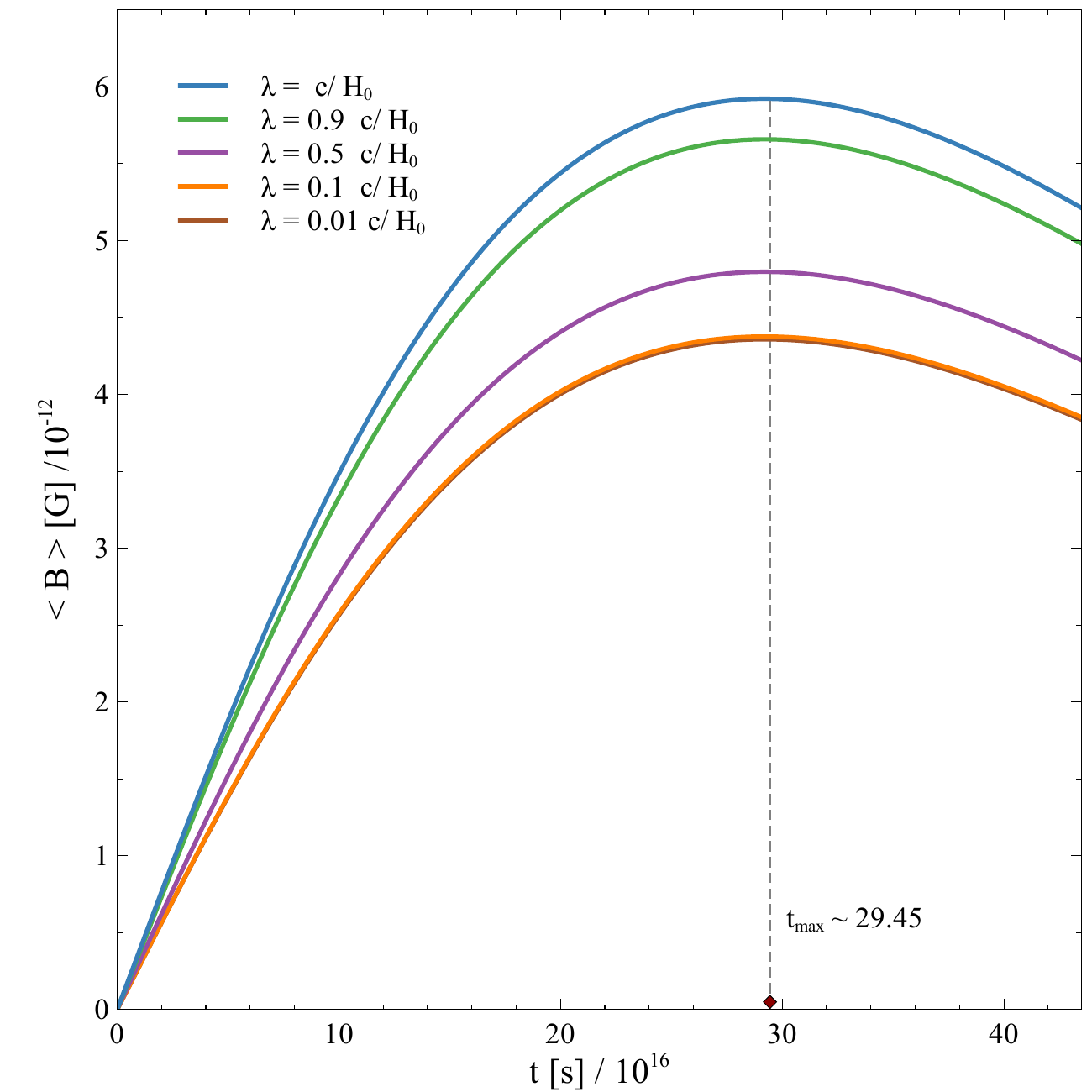}
    \caption{Temporal evolution for the expectation value of magnetic field for $n=200$, with different wavelengths sub-Hubble values $\lambda = \{0.01,\; 0.1,\; 0.5, \; 0.9,\;1\} \times c/H_0$, and $l=500$. In all cases, the magnetic field reaches its maximum at $t_{\text{max}} \sim 29.45\times 10^{16}$ s and then decreases, reaching present day values on the order of $\left<V|B(t_{f})|V\right> \simeq 10^{-12}$ G. The largest values correspond to the biggest value of $\lambda$.}
    \label{fig:BTlvar}
\end{figure}
\subsection{Expectation value of the electromagnetic fields}

To compute the expectation value of the electric and magnetic fields on the curved background spacetime governed by the Riemann manifold, we must define an extension of the nonlinear quantum algebra introduced in \cite{BMAS}, which includes the quantum differential operator $\hat{\frac{\partial}{\partial x'^{j}}}$ that acts on the operator ${}_\iota\hat\sigma^{(\alpha)\dagger}_{klm}(x^{i})$, such that, the Eq. (\ref{dife}) transforms into
\begin{equation}\label{alg}
\left[\hat{\frac{\partial}{\partial x'^{j}}}, {}_\iota\hat\sigma^{(\alpha)\dagger}_{klm}(x^{i})\right] = -\frac{i\,k'_{j}}{b^2} \,\left[{}_\iota\hat\sigma_{k'l'm'}^{(\alpha)}(x'^{i}), {}_\iota\hat\sigma^{(\alpha)\dagger}_{klm}(x^{i})\right],
\end{equation}
where ${}_\iota\hat\sigma_{klm}^{(\alpha)}(x^{i})={}_\iota \hat{B}_{klm}\,{}_\iota\sigma^{(\alpha)}(k,x^{i})$ and ${}_\iota{\hat\sigma^{(\alpha)\dagger}_{klm}}(x^{i})={}_\iota \hat B^{\dagger}_{klm}\,\left(\sigma^{(\alpha)}(k,x^{i})\right)^*$. Therefore, we obtain
\begin{equation}
  \left<V\left|\hat{\frac{\partial}{\partial x'^{j}}} \hat{\delta A}^\nu\right|V\right> = \frac{1}{b} \sum_{\iota=1}^2\sum_{l,m}\frac{(2l+1)(l-m)!}{4\pi (l+m)!}\,  k_j \, {}_\iota P^\nu(k,t)  \|{}_\iota\sigma_{klm}^{(\nu)}\|^2,
\end{equation}
where we have used the fact that
\begin{equation}\label{con}
\left<V\left| \left[{}_\iota \hat{B}_{klm},\, {}_\varsigma \hat{B}_{k'l'm'}^\dagger \right]\right|V\right> = i b\,\sqrt{2\pi}\,\frac{(2l+1)(l-m)!}{2(l+m)!} \,\delta^{(3)} (\vec{k} - \vec{k}\,')\, \delta_{\iota\varsigma} \,\delta_{ll'}\,\delta_{mm'}.
\end{equation}
Therefore, the expectation value of the electric field is
\begin{equation}
\left<V\left| \hat{\delta E}^{i}\right|V\right> = - \frac{U_0}{4\pi b} g^{ij} \sum_{\iota=1}^2 \sum_{klm} \frac{(2l+1)(l-m)!}{(l+m)!} k_j \,{}_\iota P^0(k,t)  \|{}_\iota\sigma_{klm}^{(0)}\|^2 , \hspace{1 cm} i,j=1,2,3,
\end{equation}
where the wave numbers related to spherical coordinates $(x^1,x^2,x^3) = (r,\theta,\vartheta)$, are: $k_1=k$, $k_2=l\,k$ and $k_3=m\,k$. The expectation values of the magnetic field components, Eq. (\ref{mag}), for $(\theta_*,\vartheta_*)$, given in the Eq. (\ref{fix_angles}), are
\begin{subequations}\label{exp:B}
\begin{eqnarray}
\left<V\left| \hat{\delta B}^{r}(t,r,\theta_*,\vartheta_*)\right|V\right> &=& \frac{r^2}{4\pi b}\,\left(\frac{H(t)}{H_0}\right)\, e^{\int 2H dt} \sum_{\iota=1}^2\sum_{l,m}\frac{(2l+1)(l-m)!}{(l+m)!} \nonumber \\
&\times & \left[\tilde{\alpha_1}\,l\,{}_\iota P^{\vartheta}(k,t) \|{}_\iota\sigma_{klm}^{(3)}\|^2 - m\, {}_\iota P^{\theta}(k,t)\, \|{}_\iota\sigma_{klm}^{(2)}\|^2\right], \label{r}\\ 
\left<V\left| \hat{\delta B}^{\theta}(t,r,\theta_*,\vartheta_*)\right|V\right> &=& \frac{r^2}{4\pi b} \,\left(\frac{H(t)}{H_0}\right)\,e^{\int 2H dt} \sum_{\iota=1}^2\sum_{l,m} \frac{(2l+1)(l-m)!}{(l+m)!} \nonumber \\
& \times & \left[\tilde{\alpha_1} \,k\, {}_\iota P^{\vartheta}(k,t) \|{}_\iota\sigma_{klm}^{(3)}\|^2 - \frac{m}{r^2\big(1+(H_0/c)^2\, r^2\big)}\, {}_\iota P^{r}(k,t)\, \|{}_\iota\sigma_{klm}^{(1)}\|^2\right], \label{theta} \\
\left<V\left| \hat{\delta B}^{\vartheta}(t,r,\theta_*,\vartheta_*)\right|V\right> &=& \frac{r^2}{4\pi b}\,\left(\frac{H(t)}{H_0}\right)\, e^{\int 2H dt} \sum_{\iota=1}^2\sum_{l,m} \frac{(2l+1)(l-m)!}{(l+m)!} \nonumber \\
& \times & \left[k\,{}_\iota P^{\theta}(k,t)\, \|{}_\iota\sigma_{klm}^{(2)}\|^2 - \frac{l}{r^2\big(1+(H_0/c)^2\, r^2\big)}\, {}_\iota P^{r}(k,t)\, \|{}_\iota\sigma_{klm}^{(1)}\|^2\right], \label{vartheta}
\end{eqnarray}
\end{subequations}
where we have used that $\left<V\left|\hat{\delta A}^\nu \right|V\right>=0$. Finally, the squared strength of the magnetic field can be computed as follows
\begin{equation}
    \left<V\left| \hat{\delta B}\right|V\right>^2= \left<V\left| \hat{\delta B}^{i}\right|V\right>\left<V\left| \hat{\delta B}_{i}\right|V\right>.
\end{equation}
The time evolution for the expectation value of the co-moving magnetic field, for different wavelength values $\lambda$ compared with $l=500$, is depicted in the Fig. (\ref{fig:BTlvar}). Note that, the co-moving magnetic fields reach their maximum value at $t= 29.45 \times 10^{16}$ seconds, and then smoothly decrease. However, they always take values around $10^{-12}$ G, at cosmological scales. These values agree with recent observations reported in \cite{MAGIC,ApJL950,Vovk2023}.

In Table (\ref{tab}), we present the initial and final values of the magnetic field, $\left<V|B(t_{0})|V\right>$ and $\left<V|B(t_{f})|V\right>$ respectively, for $n=200$ and various present-day wavelength values relevant for cosmological scales. Here, $t_0 = 4.35\times 10^{13}$ s corresponds to the moment when CMBR was emitted, and $t_f = 4.35\times 10^{17}$ s, represents the present-day age of the universe.

\begin{figure}
    \centering
    \includegraphics[scale=0.55]{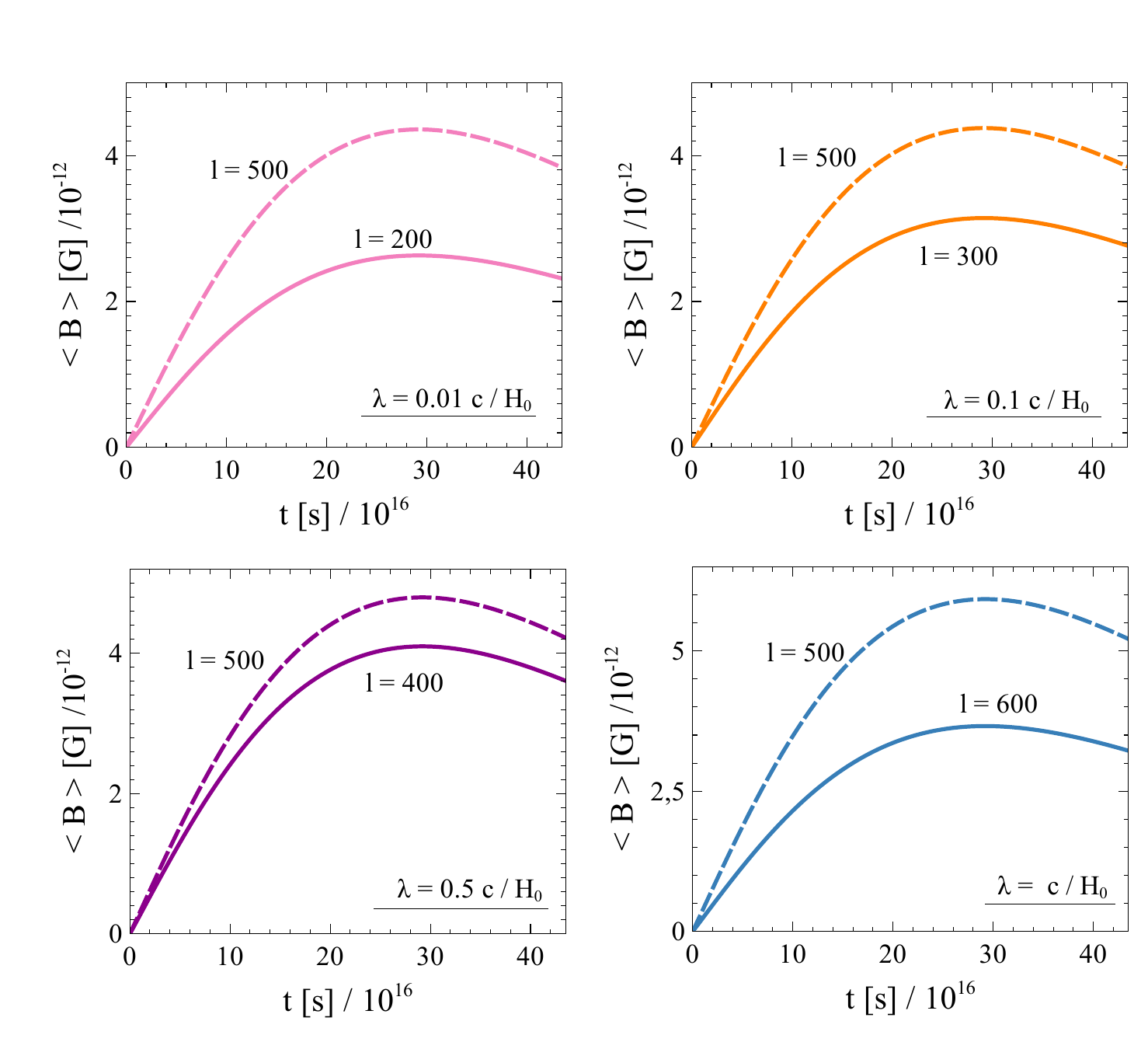}
    \caption{Temporal evolution for the expectation values for co-moving magnetic fields for $n=200$, with different wavelengths sub-Hubble values $\lambda = \{0.01,\; 0.1,\; 0.5, \; 1\} \times c/H_0 $, and $l=\{200,\; 300,\; 400, \; 600 \}$, and their comparative values with $l=500$. Notice that in all cases the co-moving magnetic fields reach their maximum at $t_*=29.45\times 10^{16}$ s, and then decrease, taking present day values of the order of $\left<V|B(t_{f})|V\right> \simeq 10^{-12}$ G. }
    \label{fig:BT2}
\end{figure}

\begin{table}[h]
    \centering
    \begin{tabular}{c c c c c}\hline\hline
         \hspace{0.4cm}$\lambda\, /10^{25}$ $[m]$\hspace{0.5cm} & \hspace{0.5cm}$k\,/10^{-25}\;\; [m^{-1}]$\hspace{0.5cm} & \hspace{0.6cm}$\left<B_0\right> \,/10^{-16}$ [G]\hspace{0.6cm} & \hspace{0.6cm}$\left<B_f\right>\,/10^{-12}$ [G]\hspace{0.6cm} & \hspace{0.6cm}$\Delta B\,/10^{-12
         }$ [G]\hspace{0.4cm} \\ [1ex]  \hline
        16.26744329 & 0.386242951 & 2.716352592 & 3.304304126 & 3.304032491\\
        13.55620274 & 0.463491541 & 4.280950842 & 5.207557948 & 5.207129853\\
        12.20058247 & 0.514990601 & 4.090227643 & 4.975552921 & 4.975143898\\
        6.778101370 & 0.926983083 & 3.467601289 & 4.218159778 & 4.217813018\\
        1.135562027 & 4.634915418 & 3.163467821 & 3.848196954 & 3.847880607\\
        0.135562027 & 46.34915418 & 3.150291818 & 3.832169020 & 3.831853990\\ \hline\hline
    \end{tabular}
    \caption{Expectation values of the magnetic field for different values of $\lambda$, $k$ and $l=500$. Here, $\left<B_{0}\right> \equiv \left<V|B(t_{0})|V\right>$ and
    $\left<B_{f}\right> \equiv \left<V|B(t_{f})|V\right>, $where $t_0 = 4.35\times 10^{13}$ s corresponds to the CMB emission, and $t_f = 4.35\times 10^{17}$ s is the age of the universe. $\Delta B \equiv |\left<B_{0}\right>-\left<B_{f}\right>|$. }
    \label{tab}
\end{table}
For completeness of the work, we also compare the magnitude of the magnetic field for different values of $l$ and fixed $\lambda$ as is indicated in the plots of Fig. \ref{fig:BT2}, it is observed that the magnitude of $\left<B\right>$ is bigger according to $l$.

\section{Conclusions}

In this work, we have explored a model of an extended general relativity, where the boundary terms are related with the perturbations of the spacetime. The extending Riemann manifold is described by the varied connection proposed into the Eq. (\ref{conec}). In this framework, we have studied the dynamics of the scalar field (which is a phantom field with negative kinetic energy density), during preinflation. We have computed the expectation value for the flux of $\hat{\Omega}^{\mu}=\hat{\sigma}^{\mu}+a\hat{\delta A}^{\mu}$ through a closed $3d$-hypersurface, which alters the Einstein equations that describes the background dynamics of the universe. This 4-vector field $\hat{\Omega}^{\mu}$ contains geometric information related to primordial gravitational and electromagnetic fluctuations. The expectation value of its flux is solely due to gravitational contributions, because we have required an electromagnetic neutral background, $\left<V\right|\nabla_{\mu}\hat{\delta A}^{\mu}\left|V\right>=0$. Therefore, the cosmological constant $\lambda_0$ is exclusively originated by the flux of $\hat{\sigma}^{\mu}= g^{\mu\nu} \hat{\partial}_{\nu} \hat{\sigma}$, as can be seen in the Eq. (\ref{aa}), which means that $\lambda_0$ alters the background dynamics described by the Einstein equations $G_{\alpha \beta} - \lambda_0 \,{g}_{\alpha \beta}=-\kappa\, T_{\alpha\beta}$, as a consequence of the rugosity of spacetime with respect to the background curvature produced by back-reaction effects studied in Subsect. (\ref{bre}).

We studied the dynamics of the  gravito-electromagnetic field during preinflation. We have obtained the general nonlinear differential Eq. (\ref{gwef}), which describes the coupled evolution of GW and electromagnetic fields. However, for simplicity, we focused on finding analytical solutions for how electromagnetic fields evolution evolve without taking into account the coupling with GW. Exploring the complex dynamics of Eq. (\ref{gwef}) would require a numerical treatment, which is beyond the scope of our current study. 
Choosing the gauge $\eta=-2b$, we obtained a decoupled dynamics for GW and electromagnetic fields in a model with variable time-scale and negative spatial curvature. In this work we have focused in the study of the electromagnetic fields. We proposed a Fourier expansion for each component of the 4-electromagnetic potential, and we found that for each pair of fixed angles $(\theta_*,\vartheta_*)$, the vector potential fluctuations $\hat {\delta A}^i$ can be written in terms of the electric potential $\hat{\Phi}$, as is reported in the Eq. (\ref{vec_pot}). The polarization modes of the electric potential are dominant with respect to those of the vector potential, as can be appreciated in the Fig.  (\ref{fig:Polarization}).

In the context of quantum electromagnetic fluctuations in preinflation, and using an extension of the quantum algebra proposed in \cite{BMAS}, we have calculated the strength of primordial co-moving magnetic fields, which were depicted in Figs. \ref{fig:BTlvar} and \ref{fig:BT2}. In all cases, the magnetic field reaches its peak at $t_{\text{max}}=29.45 \times 10^{16}$ s, and then gradually decreases. They take present-day values of the order of $\left<V|B(t_{f})|V\right> \simeq 10^{-12}$ G, which is consistent with the recent constraints reported in  \cite{MAGIC,ApJL950,Vovk2023}. It is important to notice the distinction between our results and those obtained by other models, as seen, for instance \cite{MemBell}.
In our case co-moving magnetic fields grow from small initial values during the early stages of the universe's evolution, and thereafter decrease. This significant divergence eliminates the necessity of imposing excessive large initial values upon these fields, to recover the present day observed values.

Observe that the choice of the gauge $\eta=-2b$ greatly simplifies our calculations, making it possible to make an analytical treatment of the dynamics of $\hat{\delta A}^{\mu}$.  As a prospective avenue for future research, it would be highly intriguing to explore an alternate gauge in which the coupling between $\hat{\delta A}^{\mu}$ and $\hat{\delta \Psi}_{\alpha\beta}$ can be examined. Such an investigation, particularly the exploration of the transfer of B-modes from GW to electromagnetic waves observable in the CMBR spectrum, holds notable importance. This task appears to necessitate numerical calculations for a comprehensive analysis. Along our study we have not taken into account the magnetic field interactions with plasma, which occurs on astrophysical, and not cosmological scales. Therefore our considerations are valid in a description where the magnetic fields are frozen in plasma. Accounting for the nonlinear evolution of the field changes drastically the limits of the fields measured today \cite{re1,re2}. Magnetic field's generation can be described as an injection of magnetic energy to cosmological plasma at a given scale determined by the moment of magnetic field generation \cite{re3,re6,re7}. Such fields, that are produced on small scales, can play an important role in structure formation and could provide an explanation to the apparently observed magnetic fields in the voids of the large-scale structure \cite{re4}. Because of this, it was suggested that
measuring cosmological magnetic fields in low-density environments such as filaments is much more useful than observing cluster magnetic fields to infer their possible origin \cite{re5}.

\section*{CRediT authorship contribution statement}

Daniela Magos: Ideas, Conceptualization, Calculation, Writing. Mauricio Bellini: Ideas, Conceptualization, Calculation, Writing.

\section*{Declaration of competing interest}

The authors declare the following financial interests/personal relationships which may be considered as potential competing interests:
The authors reports financial support was provided by National Scientific and Technical Research Council. The authors reports
financial support was provided by National University of Mar del Plata.

\section*{Data availability}

No data was used for the research described in the article.

\section*{Acknowledgements}

\noindent The authors acknowledge CONICET, Argentina (PIP 11220200100110CO) and UNMdP (EXA1055/22) for financial support. Daniela Magos thanks CONICET for the 2020 postdoctoral fellowship with Latin American countries.

\end{document}